\documentclass[10pt]{article}

\usepackage{latexsym,amsmath,amssymb}

\numberwithin{equation}{section}

\newcommand{\be}{\begin{equation}}
\newcommand{\ee}{\end{equation}}
\newcommand{\nn}{\nonumber}
\newcommand{\bea}{\begin{eqnarray}}
\newcommand{\eea}{\end{eqnarray}}

\newcommand{\ns}{\normalsize}

\def\a{\alpha}
\def\b{\beta}
\def\g{\gamma}

\def\d{\delta}
\def\e{\epsilon}

\def\vf{\varphi}

\def\z{\psi}
\def\k{\kappa}
\def\l{\lambda}
\def\m{\mu}
\def\n{\nu}
\def\o{\omega}
\def\p{\pi}
\def\q{\theta}

\def\r{\rho}
\def\s{\sigma}
\def\t{\tau}

\def\x{\xi}
\def\z{\zeta}

\def\L{\Lambda}
\def\O{\Omega}

\def\T{\Theta}

\def\cF{{\mathcal F}}

\def\cG{{\mathcal G}}

\def\cK{{\mathcal K}}

\def\cO{{\mathcal O}}

\def\RM{{\mathbb R}}
\def\ZM{{\mathbb Z}}
\def\PM{{\mathbb P}}
\def\CM{{\mathbb C}}

\def\au{{\underline a}}
\def\bu{{\underline b}}
\def\cu{{\underline c}}

\def\muu{{\underline \mu}}
\def\nuu{{\underline \nu}}
\def\ru{{\underline \rho}}
\def\su{{\underline \sigma}}
\def\Au{{\underline A}}

\def\bzeta{\boldsymbol{\zeta}}
\def\bxi{\boldsymbol{\xi}}

\begin{document}


\begin{titlepage}

\vspace{-3cm}

\title{
   \hfill{\ns SUSX-TH/03-004\\}
   \hfill{\ns hep-th/0305078\\}
   \vskip 1cm
   {\Large Moduli K\"ahler Potential for M-theory on a $G_2$ Manifold}\\}
   \setcounter{footnote}{0}
\author{
{\ns\large Andr\'e Lukas\footnote{email: a.lukas@sussex.ac.uk}
  \setcounter{footnote}{3}
  and Stephen Morris\footnote{email: s.morris@sussex.ac.uk}} \\[0.8em]
   {\it\ns Department of Physics and Astronomy, University of Sussex}
   \\[-0.2em]
   {\ns Falmer, Brighton BN1 9QJ, UK} \\[0.2em] }
\date{}

\maketitle

\begin{abstract}\noindent
We compute the moduli K\"ahler potential for M-theory on a compact
manifold of $G_2$ holonomy in a large radius approximation. Our method
relies on an explicit $G_2$ structure with small torsion, its periods
and the calculation of the approximate volume of the manifold.
As a verification of our result, some of the components of the
K\"ahler metric are computed directly by integration over harmonic
forms. We also discuss the modification of our result in the
presence of co-dimension four singularities and derive the gauge-kinetic
functions for the massless gauge fields that arise in this case.
\end{abstract}

\thispagestyle{empty}

\end{titlepage}


\section{Introduction}

Seven-dimensional spaces with holonomy $G_2$ provide the general
setting for relating M-theory to four-dimensional theories with $N=1$
supersymmetry. It has been known for some time~\cite{Witten:1983ux}
that 11-dimensional supergravity on smooth seven-dimensional manifolds
reduces to a non-chiral theory in four dimensions. More specifically,
for smooth manifolds of holonomy $G_2$ the four-dimensional spectrum
consists of Abelian gauge multiplets, which descend from the
three-form of 11-dimensional supergravity, and uncharged chiral
multiplets that contain the metric moduli of the $G_2$ manifold and
associated axions~\cite{Papadopoulos:1995da}.

The situation changes if the $G_2$ space acquires singularities.
Specifically, singularities of co-dimension four lead to
non-Abelian gauge multiplets and co-dimension seven singularities
to chiral matter (possibly charged under these gauge
multiplets)~\cite{Acharya:1998pm}--\cite{Atiyah:2001qf}. These
features make M-theory on singular $G_2$ spaces an interesting
framework for M-theory ``particle phenomenology'' and have
triggered much activity in the subject
recently~\cite{Acharya:2000ps}--\cite{Acharya:2002kv}.

\vspace{0.7cm}

A more detailed analysis of the phenomenology of such models requires
explicit knowledge of the four-dimensional effective theory. It is the
main purpose of this paper to work out some of its
features. Concretely, we will compute the four-dimensional moduli
K\"ahler potential and the gauge kinetic functions obtained from
M-theory on a $G_2$ manifold. This requires working with compact $G_2$
manifolds rather that with the non-compact
examples~\cite{Gibbons:er,Chong:2002yh} that have been widely used in
recent work. The moduli K\"ahler potential is obviously relevant to a
number of problems in this context, for example to the study of
supersymmetry breaking and the cosmological dynamics of moduli, to
name only two.

For reasons of simplicity, we will initially consider a smooth $G_2$
manifold $X$ and later allow for co-dimension four singularities
leading to non-Abelian gauge fields. In this paper, we will not
attempt to include co-dimension seven singularities. For concreteness,
we will focus on the specific compact $G_2$ manifold constructed by
Joyce in Ref.~\cite{joyce1}. However, our method applies to a large class
of compact $G_2$ manifolds constructed in a similar
fashion~\cite{joyce2,joyceb}.

\vspace{0.7cm}

For some classes of internal manifolds a general and explicit formula
for the moduli K\"ahler potential (at least at tree level) can be given
in terms of certain topological data. For example, in the case of the
K\"ahler moduli space of Calabi-Yau three folds the moduli K\"ahler
potential is determined by a cubic polynomial with coefficients given
by the triple intersection numbers of the Calabi-Yau space in
question~\cite{Strominger:ks,Candelas:1990pi}.  At the heart of this
result is a quasi topological formula for the Hodge dual of two-forms
on the Calabi-Yau space. Unfortunately, an analogous formula for
three-forms on $G_2$ manifolds does not seem to exist. Therefore,
while abstract formulae for the moduli K\"ahler metric in terms of
harmonic three-forms on the $G_2$ manifold, and for the K\"ahler
potential in terms of the volume of the $G_2$ manifold, are
known~\cite{Beasley:2002db}, these expressions cannot be evaluated
generically for all $G_2$ manifolds in the way they can for the
Calabi-Yau K\"ahler moduli space. Our approach will, therefore, be to
focus on the particular $G_2$ manifold of Ref.~\cite{joyce1} and
explicitly construct all the objects required. Concretely, on a
specific $G_2$ manifold $X$, we will construct a family of $G_2$
structures $\vf$ with small torsion, following Ref.~\cite{joyce1}, and
determine the associated family of ``almost Ricci-flat'' metrics
$g$. By computing the periods of $\vf$ and the volume as measured by
the metrics $g$ we are able to compute the moduli K\"ahler potential
in a controlled approximation.

\vspace{0.7cm}

Let us now summarise the main result of this paper. We have
computed the four-dimensional moduli K\"ahler potential for
M-theory on the compact $G_2$ manifold $X$ constructed in
Ref.~\cite{joyce1}. The main features of this manifold are as
follows. The starting point of the construction is the seven-torus
$T^7=\RM^7/\ZM^7$ with coordinates $x^A$, where $A,B,\dots =
1,\dots ,7$. This torus is divided by three $\ZM_2$ symmetries,
generated by $\a$, $\b$ and $\g$, whose precise action on the
coordinates $x^A$ is given in Eqs.~\eqref{a}, \eqref{b} and
\eqref{g}. The resulting orbifold has $12$ co-dimension four fixed
points, four associated with each of the three $\ZM^2$ symmetries.
We will label these fixed points by a pair $(\t ,n)$, where $\t = \a
,\b ,\g$ indicates the type of the fixed point (that is, under which
of the $\ZM^2$ symmetries it remains inert) and $n=1,2,3,4$ labels
the fixed points of the same type. The manifold $X$ is then obtained
by blowing up each of these $12$ points using Eguchi-Hanson
spaces. We will refer to these regions of $X$ as ``blow-ups'' and
to the surrounding torus-like region as ``bulk''.

The metric moduli of this space can be organised into bulk moduli
and moduli associated with the blow-ups. It turns out that the
only bulk parameters that survive the orbifolding are the seven radii
of the torus, producing seven corresponding moduli $a^A$. The
precise relation between $a^A$ and the geometrical radii of the
torus is given in Eq.~\eqref{aA}. Further, each Eguchi-Hanson
blow-up comes with three additional moduli, one being the radius
of the blow-up, the other two describing the orientation of the
Eguchi-Hanson space relative to the bulk. We denote these moduli
by $a^{(\t ,n,a)}$, where $a,b,\dots = 1,2,3$, and their relation
to the underlying geometrical parameters is given in
Eq.~\eqref{ai}. In total, we therefore have $7+3\cdot 12=43$
metric moduli. They pair up with $43$ axions, which descend from
the three-form of 11-dimensional supergravity, to form $N=1$ chiral
superfields. We denote these chiral superfields by $T^A$ and
$U^{(\t ,n,a)}$, where ${\rm Re}(T^A)=a^A$ and ${\rm Re}(U^{(\t
,n,a)})=a^{(\t ,n,a)}$. For the K\"ahler potential $K$ of these
fields we find
\begin{eqnarray}
 K &=& -\sum_{A=1}^7\ln (T^A+\bar{T}^A)-3\ln\left[1-\frac{8}{3}
     \sum_{\t ,n,a}\frac{(U^{(\t ,n,a)}+\bar{U}^{(\t ,n,a)})^2}
     {(T^{A(\t ,a)}+\bar{T}^{A(\t, a)})(T^{B(\t ,a)}+\bar{T}^{B(\t ,a)})}
     \right] + c \label{K}\\
   &\simeq&-\sum_{A=1}^7\ln (T^A+\bar{T}^A)+8\sum_{\t ,n,a}
     \frac{(U^{(\t ,n,a)}+\bar{U}^{(\t ,n,a)})^2}
     {(T^{A(\t ,a)}+\bar{T}^{A(\t ,a)})(T^{B(\t ,a)}+\bar{T}^{B(\t ,a)})}+c\; .
     \label{Kapp}
\end{eqnarray}
Here $A(\t ,a)$ and $B(\t ,a)$ are two specific indices of
$1,\dots ,7$ that determine the two bulk moduli by which the
blow-up moduli $U^{(\t ,n,a)}$ associated with the fixed point
$(\t ,n)$ are divided. The specific values of these index
functions, which are directly linked to the structure of the
orbifolding, are given in Table 1. The above K\"ahler potential
constitutes an approximate results for two reasons. Firstly, as
usual one has to require all moduli to be sufficiently large for
the supergravity approximation to be valid.  Secondly, the blow-up
moduli $U^{(\t ,n,a)}$ have to be small compared to the bulk
moduli $T^A$, so that the expected corrections of quartic and
higher order in $U/T$ can be neglected. We have expanded the
logarithm in the second line~\eqref{Kapp} to indicate that our
result is accurate to leading, quadratic order in $U/T$.  The
constant $c$ is irrelevant for the moduli kinetic terms associated
with $K$ but it does play a role in the presence of a non-trivial
superpotential, such as that based on flux proposed in
\cite{Gukov:1999gr,Acharya:2000ps,Beasley:2002db}. The value of
$c$ for our normalisation of the fields is $6\ln(8\pi)+\ln(2)$
(see Eq.~\eqref{c}). Note that one cannot consistently work with a
universal bulk modulus while varying the blow-up moduli
independently for different values of $(\t ,a)$. From
Eq.~\eqref{K}, each pair $(\t ,a)$ is sensitive to two particular
bulk moduli so that such a non-universal evolution of the blow-up
moduli almost inevitably leads to anisotropy in the bulk
evolution.

\vspace{0.7cm}

In addition to the moduli chiral superfields, the four-dimensional
effective theory also contains Abelian vector multiplets. For our
specific $G_2$ manifold we have $12$ such multiplets, one for each
blow-up. The gauge-kinetic functions for these multiplets depend
on the type $\t$ of the blow-up only and are given by
\begin{equation}
 f_{(\t )}=\left\{\begin{array}{lll}
   T^7&{\rm for}&\t =\a\\
   T^6&{\rm for}&\t =\b\\
   T^5&{\rm for}&\t =\g
 \end{array}\right.\; . \label{f0}
\end{equation}
What happens if any of the blow-ups collapse to an orbifold
singularity? One expects the moduli K\"ahler potential to still be
of the form~\eqref{K} but with all terms corresponding to
singularities dropped. This amounts to blowing down all
singularities $(\t ,a)$ via $U^{(\t ,n,a)}\rightarrow 0$. For each
such singularity the gauge group enhances from ${\rm U}(1)$ to ${\rm SU}(2)$.
The gauge kinetic functions $f$ of the associated ${\rm SU}(2)$
gauge multiplets are still given by the same expressions~\eqref{f0}
as in the Abelian case.

\vspace{0.7cm}

The plan of the paper is as follows. In the next section, we will
review some necessary facts about M-theory on $G_2$ manifolds and
outline our strategy to compute the K\"ahler potential~\eqref{K}. In
Section 3, we construct the manifold $X$ and a basis of its third
homology, following Ref.~\cite{joyce1}. A family of $G_2$ structures
$\vf$ and the associated family of metrics $g$ on $X$ is presented in
Section 4. Using these ingredients, in Section 5, we compute the
periods of $\vf$ and the volume of $X$ in a controlled
approximation. Combining these results, we finally obtain the moduli
K\"ahler potential. In Section 6, as a check for our result, we verify
that some of the components of the associated K\"ahler metric can be
reproduced by a direct calculation using some of the harmonic forms on
$X$. Finally, in Section 7, we discuss gauge-kinetic functions
and the effect of co-dimension four singularities.

To keep the main text more readable we have collected many of the
technical details in three Appendices; on Eguchi-Hanson spaces,
smoothed Eguchi-Hanson spaces and $G_2$ structures.

\vspace{0.7cm}

Let us summarise the main conventions we will use in this paper.
We denote seven-dimensional coordinates by ${\bf x}=(x^A)$ with associated
indices $A,B,\dots = 1,\dots ,7$. We will frequently need to
split these seven coordinates into a group of four coordinates,
denoted by $\bzeta = (\z^\m )$, with indices $\m ,\n ,\dots = 0,1,2,3$
and a complementary group of three coordinates denoted by $\bxi = (\x^a)$
with indices $a,b,\dots = 1,2,3$. We will use four-dimensional coordinates
${\bf z}=(x,y,z,t)=(z^\m )$ on the Eguchi-Hanson spaces. Tangent space
indices are denoted by an underlined version of their curved counterparts.

As mentioned above, the various Eguchi-Hanson blow-ups are
labelled by a pair $(\t ,n)$, where $\t = \a ,\b ,\g$ indicates the
type and $n=1,2,3,4$ numbers the four blow-ups of each type. In
dealing with the specific $G_2$ manifold we consider, we will find
it practical to work in the ``upstairs" picture for the orbifold,
that is, we will consider the full torus $T^7$ rather than a
fundamental domain.  In this picture, there exist four equivalent
copies of each blow-up. We will label these degenerate blow-ups by
adding an additional index $d=1,2,3,4$, that is, we use a triple
$(\t ,n,d)$ of indices. For notational simplicity, we will
frequently abbreviate this triple by $(i)=(\t ,n,d)$. Generic moduli of
the $G_2$ manifold will be indexed by $I,J,\dots$.


\section{M-theory on $G_2$ manifolds}

Let us start by considering a seven-dimensional real, compact
manifold $X$ with $G_2$ holonomy and coordinates ${\bf x}=(x^A)$.
Such a manifold is equipped with a $G_2$ structure $\vf$, that is,
a smooth three-form which is isomorphic to the ``flat" $G_2$
structure
\begin{eqnarray}
 \bar{\vf} &=& dx^1\wedge dx^2\wedge dx^7 + dx^1\wedge dx^3\wedge dx^6
            + dx^1\wedge dx^4\wedge dx^5\nn \\
            && + dx^2\wedge dx^3\wedge dx^5 + dx^4\wedge dx^2\wedge dx^6
               + dx^3\wedge dx^4\wedge dx^7 + dx^5\wedge dx^6\wedge dx^7\; .
 \label{phib0}
\end{eqnarray}
By way of this isomorphism, a $G_2$ structure induces a Riemannian
metric $g$ on $X$ that can be explicitly computed from $\vf$ using
Eqs.~\eqref{gammadef} and \eqref{gdef}. We call $g$
the metric associated with the $G_2$ structure $\vf$. The
associated metric can be used to define the map $\T$ on $G_2$
structures by $\T (\vf )=\star\vf$, where the Hodge-star is with
respect to the associated metric of the argument $\vf$ of $\T$.
The additional $\vf$ dependence hidden in the Hodge star makes
this map highly non-linear.  On a manifold with holonomy $G_2$
there exists a torsion-free $G_2$ structure, that is, a $G_2$
structure satisfying $d\vf =d\T(\vf )=0$, or, equivalently, for
compact $X$, a $G_2$ structure which is harmonic with respect to
its associated metric. We will denote such a torsion-free $G_2$
structure $\tilde{\vf}$. Its associated metric $\tilde{g}$ is a
Ricci-flat metric on $X$.

\vspace{0.7cm}

The Ricci-flat deformations of the metric $\tilde{g}$ can be described
by the torsion-free deformations of the $G_2$ structure $\tilde{\vf}$
and, hence, by the third cohomology $H^3(X,\RM )$. Consequently,
the number of independent metric moduli is given by the third Betti
number $b^3(X)$. To define these moduli more explicitly, we first
introduce an integral basis $\{ C^I\}$ of three-cycles,
where $I,J,\dots = 1,\dots ,b^3(X)$, and an associated dual
basis $\{\Phi_I\}$ of harmonic three-forms. Duality implies
that
\begin{equation}
 \int_{C^I}\Phi_J = \d_J^I\; . \label{dual}
\end{equation}
A torsion-free $G_2$ structure $\tilde{\vf}$ can then be expanded
as
\begin{equation}
 \tilde{\vf} = \sum_I a^I\Phi_I\; ,
\end{equation}
where the expansion coefficients $a^I$ represent the metric moduli in
question. From Eq.~\eqref{dual}, these moduli $a^I$ can be computed
in terms of certain underlying geometrical parameters (on which
a family of $G_2$ structures $\tilde{\vf}$ depends) by performing
the period integrals
\begin{equation}
 a^I =\int_{C^I}\tilde{\vf}\; . \label{periods}
\end{equation}

\vspace{0.7cm}

Let us also introduce an integral basis $\{D^I\}$ of two-cycles,
where $I,J,\dots = 1,\dots ,b^2(X)$ and a dual basis $\{\o_I\}$ of
two-forms satisfying
\begin{equation}
 \int_{D^I}\o_J = \d_J^I\; .
\end{equation}
Then, the three-form field $C$ of 11-dimensional supergravity
can be expanded in terms of the basis $\{\Phi_I\}$ and $\{\o_I\}$ as
\begin{equation}
 C = \nu^I\Phi_I+A^I\wedge \o_I\; . \label{C}
\end{equation}
The expansion coefficients $\nu^I$ represent $b^3(X)$ axionic fields
in the four-dimensional effective theory, while the Abelian gauge
fields $A^I$, with field strengths $F^I$, are part of $b^2(X)$ Abelian
vector multiplets. The $\nu^I$ pair up with the metric moduli $a^I$ to
form the bosonic parts of $b^3(X)$ four-dimensional chiral superfields
\begin{equation}
 T^I=a^I+i\n^I\; .
\end{equation}
It is the K\"ahler potential for these fields $T^I$ we wish to compute
explicitly. A general formula for the associated K\"ahler metric
$K_{I\bar{J}}$ is given by~\cite{Beasley:2002db}
\begin{equation}
 K_{I\bar{J}} = \frac{1}{4V}\int_X\Phi_I\wedge\star\Phi_J\; ,
 \label{KIJ}
\end{equation}
where the Hodge star is defined with respect to the Ricci-flat
metric $\tilde{g}$ and $V$ is the volume
\begin{equation}
 V = \int_X\sqrt{{\rm det}(\tilde{g})} \label{V}
\end{equation}
of $X$. This formula is most easily proven by reducing the kinetic
term for $C$ in the action of 11-dimensional supergravity by
inserting the expansion~\eqref{C}. With some more effort, it can
also be derived by reducing the 11-dimensional Einstein-Hilbert
term~\cite{Gutowski:2002dk}.

Using general properties of $G_2$ manifolds, it was shown in
Ref.~\cite{Beasley:2002db} that the K\"ahler metric~\eqref{KIJ} descends from
the K\"ahler potential
\begin{equation}
 K = -3\ln\left(\frac{V}{2\p^2}\right) \label{Kdef}
\end{equation}
with the volume $V$ defined in~\eqref{V}. This means that the
associated K\"ahler metric
\begin{equation}
 K_{I\bar{J}}=\frac{\partial^2 K}{\partial T^I\partial\bar{T}^J}
 \label{KIJ1}
\end{equation}
must coincide with the one obtained directly from Eq.~\eqref{KIJ}.
It can also be shown that the first derivatives of the K\"ahler
potential
\begin{equation}
 K_I=\partial K/\partial T^I
\end{equation}
can be directly computed from the harmonic forms $\Phi_I$ using
\begin{equation}
 K_I=\frac{1}{2V}\int_X\Phi_I\wedge\T (\tilde{\vf})\; .\label{KI}
\end{equation}
These facts will later provide us with a useful check of our explicit
result for the K\"ahler potential.

It is clear from Eq.~\eqref{Kdef} that the K\"ahler potential (and the
K\"ahler metric) only depends on the metric moduli $a^I$ but not on
the axions $\n^I$. In terms of superfields, this means that $K$ is a
function of the real parts $T^I+\bar{T}^I$ only.

\vspace{0.7cm}

Reduction of the Chern-Simons term of 11-dimensional supergravity by
inserting the gauge field part of \eqref{C} leads to the
four-dimensional term~\cite{Papadopoulos:1995da}
\begin{equation}
 \int_{M_4} c_{IJK}\,\n^I F^J\wedge F^K\; ,
\end{equation}
where the coefficients $c_{IJK}$ are given by
\begin{equation}
 c_{IJK} \propto \int_X\Phi_I\wedge\o_J\wedge\o_K\; . \label{cIJK}
\end{equation}
This implies that the gauge-kinetic function $f_{JK}$, which couples
$F^J$ and $F^K$, is of the form
\begin{equation}
 f_{JK} \propto \sum_I T^Ic_{IJK}\; . \label{fJK}
\end{equation}

\vspace{0.7cm}

Let us summarise our strategy to compute $K$. First, we will
explicitly construct a specific manifold $X$, a basis $\{ C^I\}$
of its third homology and a family of $G_2$ structures on $X$
that depend on a number of geometrical parameters, such as radii.
We denote these geometrical parameters collectively by $R^I$. This
part of the construction closely follows Ref.~\cite{joyce1}. Using
the relations, ~\eqref{gammadef} and \eqref{gdef}, between $G_2$
structures and metrics, we then compute the family of associated
metrics. These metrics can be used, via Eqs.~\eqref{V} and
\eqref{Kdef}, to compute the volume and the K\"ahler potential as
a function of $R^I$. Next, we evaluate the periods to obtain the
moduli $a^I$ as a function of $R^I$, which, in turn, allows us to
rewrite the K\"ahler potential in terms of the proper superfields
$T^I$.

Finally, by computing the associated K\"ahler metric from Eq.~\eqref{KIJ1}
we obtain predictions for the integrals~\eqref{KIJ}. As a check
of our result, we will determine some of the harmonic forms
$\Phi_I$ on $X$ and verify these predictions by computing some
of the integrals~\eqref{KIJ} directly.

Ideally, one would like to carry out this program using a family
of torsion-free $G_2$ structures $\tilde{\vf}$. However, such
torsion-free structures are not explicitly known on compact $G_2$
manifolds. Rather, the construction of Ref.~\cite{joyce1} involves
explicitly writing down $G_2$ structures $\vf$ with small torsion
and proving the existence of ``nearby" torsion-free $G_2$
structures. We will perform our computation using these small
torsion $G_2$ structures and show that this allows one to compute
the K\"ahler potential in a controlled approximation.


\section{Construction of the manifold}

We will now review the construction of the manifold $X$. The starting
point is the seven-dimensional standard torus $T^7=\RM^7/\ZM^7$ with
coordinates ${\bf x}=(x^A)$ and the associated orbifold
$O=T^7/\ZM_2^3$. Here, the three $\ZM_2$ symmetries are generated by
$\a$, $\b$ and $\g$ acting on the torus coordinates as
\begin{eqnarray}
 \a ((x^1,\dots ,x^7)) &=& \left( -x^1,-x^2,-x^3,-x^4,x^5,x^6,x^7\right)
 \label{a}\\
 \b ((x^1,\dots ,x^7)) &=& \left( -x^1,\frac{1}{2}-x^2,x^3,x^4,-x^5,-x^6
                           ,x^7\right) \label{b}\\
 \g ((x^1,\dots ,x^7)) &=& \left( \frac{1}{2}-x^1,x^2,\frac{1}{2}-x^3,x^4,
                           -x^5,x^6,-x^7\right)\; . \label{g}
\end{eqnarray}
Let us discuss the fixed loci of this orbifold, which are of
co-dimension four. Inspection of the above generators shows that
each of the three $\ZM_2$ symmetries leaves $16$ three-tori,
$T^3=\RM^3/\ZM^3$, invariant. Some of these are mapped into each
other by the remaining symmetries, leaving four inequivalent fixed
tori for each $\ZM_2$ and twelve in total. However, since we are
going to use the ``upstairs'' picture, that is consider the full
torus $T^7$ rather than a fundamental domain of the orbifold, we
will effectively work with $48$ fixed tori, which have to be
identified in groups of four. These $48$ three-tori are denoted by
$T_{(i)}$ where the label $(i)$ is split as $(i)=(\t ,n,d)$
whenever necessary. Here, the type  $\t = \a ,\b ,\g$ indicates
under which of the three $\ZM_2$ symmetries the three-torus
remains invariant, $n=1,2,3,4$ labels the four inequivalent tori
for a given type and $d=1,2,3,4$ counts the four-fold degeneracy
due to the upstairs picture.

To describe these fixed tori in more detail, it is useful to split,
for each symmetry type $\t =\a ,\b ,\g$, the coordinates ${\bf x}=(x^A)$
into a group ${\bzeta_\t}=(\z_\t^\m )$ of four coordinates which
transforms non-trivially under the $\ZM_2$ and a complementary
group ${\bxi_\t}=(\xi^a_\t )$ of three coordinates which transform
trivially. For the three symmetries we have
\begin{itemize}
 \item $\ZM_2$ generated by $\a$
 \begin{eqnarray}
   \bzeta_{\a} &=& (x^1,x^2,x^2,x^4)\; ,\qquad \bxi_{(\a )}=(x^7,x^6,x^5)
   \label{xa}\\
   T_{(\a ,n,d)} &=&\left\{\zeta_{(\a )}^\m\in\left\{ 0,\frac{1}{2}\right\}
   \right\}
 \end{eqnarray}
 \item $\ZM_2$ generated by $\b$
 \begin{eqnarray}
   \bzeta_{\b} &=& (x^5,x^6,x^1,x^2)\; ,\qquad\bxi_{(\b )}=(x^7,x^4,x^3)
   \label{xb}\\
   T_{(\b ,n,d)} &=&\left\{\z_{(\b)}^0,\z_{(\b )}^1,\z_{(\b )}^2
                 \in\left\{ 0,\frac{1}{2}\right\}\mbox{ and }\z_{(\b )}^3\in
                 \left\{\frac{1}{4},\frac{3}{4}\right\}\right\}
 \end{eqnarray}
 \item $\ZM_2$ generated by $\g$
 \begin{eqnarray}
   \bzeta_{\g} &=& (x^7,x^1,x^3,x^5)\; ,\qquad\bxi_{(\g )}=(x^2,x^4,x^6)
   \label{xc}\\
   T_{(\g ,n,d)} &=&\left\{\z_{(\g )}^0,\z_{(\g )}^3\in\left\{ 0,\frac{1}{2}
                    \right\}\mbox{ and }\z_{(\b )}^1,\z_{(\b )}^2\in
                     \left\{\frac{1}{4},\frac{3}{4}\right\}\right\}\; .
 \end{eqnarray}
\end{itemize}

The next step in the construction of $X$ is to remove, for each fixed
point $(i)$, a four-dimensional ball centred around this fixed point
times the associated fixed three-torus $T_{(i)}$. We will refer to the
remaining parts of the torus $T^7$ as the bulk $B$. The holes in $B$
are then replaced by $F_{(i)}\equiv U\times T_{(i)}$, where $U$ is the
blow-up of $\CM^2/\ZM_2$ as discussed in Appendix~\ref{EH}. Hence, the
manifold $X$ consists of a bulk chart $B$ and charts $F_{(i)}$ for
each of the fixed points of the underlying orbifold. We denote the
coordinates in $F_{(i)}$ by ${\bf z}_{(i)}=(z_{(i)}^\m )$ and allow
for a general linear transformation
\begin{equation}
 \bzeta_{(i)} = G_{(i)}{\bf z}_{(i)}\; , \label{trans}
\end{equation}
where $G_{(i)}\in{\rm Gl}(4)$, as a transition function in the
four-dimensional part of the overlap between $B$ and $F_{(i)}$.

\vspace{0.4cm}

Let us now present a basis of three-cycles. First of all, we have
seven bulk three-cycles $C^A\subset B$ which correspond to the seven
terms in the flat $G_2$ structure~\eqref{phib0}. They can be defined
by setting four of the coordinates $x^A$ to constants (chosen so there
is no intersection with the blow-ups $F_{(i)}$) and adding on the
seven images identified under $\ZM_2^3$. For concreteness, we define
the cycle $C^A$ by setting the four coordinates on which the $A^{th}$
term in~\eqref{phib0} does not depend to constants, that is, for
example
\begin{equation}
 C^1 = \left\{ x^3,x^4,x^5,x^6=\mbox{const}\right\}\cup\left\{
       \mbox{seven copies under }\ZM_2^3\right\} \label{CA}
\end{equation}
and similarly for the other cycles. We move on to the three-cycles
localised in the blow-up $F_{(i)}=U\times T_{(i)}$. Let us first
denote the exceptional divisor (see Appendix \ref{EH} for a definition) in $U$ by $E_{(i)}$ and the one-cycle
along the coordinate direction $\xi_{(\t )}^a$ in $T_{(i)}$ by
$L_{(i)}^a$. Then, for each fixed point $(\t ,n)$, we can define
three three-cycles
\begin{equation}
 C^{(\t ,n,a)} = \cup_{d=1}^4E_{(\t ,n,d)}\times L_{(\t ,n,d)}^a
               \subset \cup_{d=1}^4 F_{(\t ,n, d)} \label{Ctna}
\end{equation}
The union over $d$ is a result of us working in the upstairs
picture and it accounts for the images identified by the
orbifolding. The collection $\{ C^A,C^{(\t ,n,a)}\}$ of cycles
then provides a basis for $H_3(X,\ZM )$. In total there are
$7+12\cdot 3=43$ cycles and we have $b^3(X)=43$.

Two-cycles can, in general originate from the bulk and the
blow-ups. Inspection of the orbifold action~\eqref{a}--\eqref{g} shows
that there are no $\ZM_2^3$ invariant forms $dx^A\wedge dx^B$ and,
hence, no bulk two-cycles. Each blow-up comes with a single two-cycle
given by the exceptional divisor. We, therefore, have $12$ two-cycles
which we denote by $\{D^I\}=\{D^{(\t ,n)}\}$. Explicitly, they are given
by
\begin{equation}
 D^{(\t ,n)} = \cup_{d=1}^4E_{(\t ,n,d)}\; .\label{DI}
\end{equation}
As a result we have $b^2(X)=12$.


\section{$G_2$ structures and associated metrics}

Let us now explain how to construct a family of $G_2$ structures $\vf$
on the manifold $X$ of the previous section. In doing so, we need
to include the full dependence on all $43$ moduli.

We begin with the $G_2$ structures on the bulk $B$. It is easy to
see that, for a constant metric on the torus $T^7$, only the
diagonal components survive the orbifolding by~\eqref{a},
\eqref{b} and \eqref{g}. These diagonal components are the seven
radii $R^A$ of the torus. We easily obtain the bulk $G_2$
structure from the flat $G_2$ structure~\eqref{phib0} by rescaling
$x^A\rightarrow R^Ax^A$, leading to
\begin{eqnarray}
 \vf &=& R_1R_2R_7dx^1\wedge dx^2\wedge dx^7 +
               R_1R_3R_6dx^1\wedge dx^3\wedge dx^6 +
               R_1R_4R_5dx^1\wedge dx^4\wedge dx^5\nn \\
            && + R_2R_3R_5dx^2\wedge dx^3\wedge dx^5
               + R_4R_2R_6dx^4\wedge dx^2\wedge dx^6
               + R_3R_4R_7dx^3\wedge dx^4\wedge dx^7\nn \\
            && + R_5R_6R_7dx^5\wedge dx^6\wedge dx^7\; .
 \label{phiB}
\end{eqnarray}
We will also sometimes find it convenient to work with the
rescaled coordinates $\bar{x}^A=R^Ax^A$. Obviously, this
part of the $G_2$ structure is torsion-free.

To deal with the blow-ups we split the seven radii $R^A$ into a
group $R_{(\t )}^\m$ of four and a complementary group $R_{(\t
)}^a$ of three depending on the type $(\t )$ of the blow-up and in
analogy with the split~\eqref{xa}, \eqref{xb} and \eqref{xc} of
the coordinates. For a given blow-up $F_{(i)}=U\times T_{(i)}$, we
then introduce rescaled coordinates $\bar{z}_{(i)}^\m =R_{(\t
)}^\m z_{(i)}^\m$ on $U$ and $\bar{\xi}_{(\t )}=R_{(\t
)}^a\xi_{(\t )}^a$ on $T_{(i)}$. With these definitions, the $G_2$
structure on $F_{(i)}$ can be written as
\begin{equation}
 \vf = \sum_{a,b}w^a(\bar{\bf z}_{(i)},\r_{(i)})\wedge d\bar{\xi}_{(\t )}^b
       \cO_{(i)ab}-d\bar{\xi}_{(\t )}^1\wedge d\bar{\xi}_{(\t )}^2
       \wedge d\bar{\xi}_{(\t )}^3
 \label{phii}
\end{equation}
where
\begin{eqnarray}
 w_1({\bf z},\r ) &=& \frac{\cF '(u,\r )}{2}du\wedge\s_1+\cF (u,\r )
                      \s_2\wedge \s_3 \label{ww1}\\
 w_2({\bf z}) &=& udu\wedge\s_2+u^2\s_3\wedge\s_1\label{ww2} \\
 w_3({\bf z}) &=& udu\wedge\s_3+u^2\s_1\wedge\s_2\label{ww3}
\end{eqnarray}
and $u=|{\bf z}|$. Further, the function $\cF$ has been computed
to order $\r^6$, in Appendix~\ref{SEH} (see Eq.~\eqref{SF}) and
is given by
\begin{equation}
 \cF (u,\r ) = u^2+\frac{1}{2u^2}(\e^2-u\e\e ')\r^4 +\cO (\r^8)\; ,
 \label{Fdef}
\end{equation}
Here $\e$ is an interpolating function with $\e (u)=1$ for $u\leq u_0$
and $\e (u)=0$ for $u\geq u_1$; the two fixed radii $u_0$, $u_1$
satisfying $\r\ll u_0 <u_1$. The one-forms $\s^a$ are the Maurer-Cartan
forms on $S^3\simeq {\rm SU}(2)$ defined in Appendix~\ref{EH}.
Finally, $\cO_{(i)}\in {\rm SO}(3)$.

The technicalities associated with the $G_2$ structure~\eqref{phii}
are explained in detail in the Appendices, particularly in
Appendix~\ref{G2structures}. Here, we will discuss its
interpretation. The two-form $w_1$ is the K\"ahler form associated
with a ``smoothed'' Eguchi-Hanson space on the blow-up $U$ and
interpolates, via the function $\e$, between Eguchi-Hanson space for
$u< u_0$ and flat space for $u> u_1$. The parameter $\r_{(i)}$ can be
interpreted as the radius of the blow-up $(i)$.  On the Eguchi-Hanson
space ($u< u_0$), and flat space ($u>u_1$), the two-forms $w_2$ and
$w_3$ constitute the two other K\"ahler forms that are expected on
these hyperk\"ahler spaces. Since the forms $w^a$ are closed
everywhere, the same is true for the $G_2$ structure~\eqref{phii}. Due
to the hyperk\"ahler structures at $u<u_0$ and $u>u_1$ it is even
torsion-free in these regions. It departs from non-zero torsion in the
``collar'' region $u\in [u_0,u_1]$ where it interpolates between a
non-trivial torsion-free $G_2$ structure at small radius and the flat
$G_2$ structure at large radius. However, it can be shown that the
torsion $d\T (\vf )$ in these collar regions is proportional to
$\r_{(i)}$ and derivatives of $\e$.  Hence, for sufficiently small
blow-ups, $\r_{(i)}\ll 1$, and a ``smooth'' interpolation, the
deviation from a torsion-free $G_2$ structure is small. This fact will
be used to show that the K\"ahler potential can be reliably computed,
to leading order in $\r_{(i)}$, using the above $G_2$ structure.

Let us explicitly verify that the $G_2$ structure~\eqref{phii}
indeed exactly matches the bulk $G_2$ structure~\eqref{phiB}
for $u>u_1$. We know from Appendix~\ref{SEH} that, in this
region, the forms $w^a$ coincide with the three K\"ahler forms
$\bar{w}^a$ on flat space, defined in Eq.~\eqref{wb}. Hence,
for $u>u_1$ we have
\begin{equation}
 \vf =  \sum_{a,b}w^a(\bar{\bf z}_{(i)})\wedge d\bar{\xi}_{(\t )}^b
       \cO_{(i)ab}-d\bar{\xi}_{(\t )}^1\wedge d\bar{\xi}_{(\t )}^2
       \wedge d\bar{\xi}_{(\t )}^3
\end{equation}
Let us now relate the coordinates $\bar{\bf z}_{(i)}$ on $F_{(i)}$
and the bulk coordinates $\bar{\bzeta}_{(i)}$ by an ${\rm SO}(4)$
rotation $\L_{(i)}$, that is $\bar{\bf
z}_{(i)}=\L_{(i)}\bar{\bzeta}_{(i)}$, by choosing the transition
function in Eq.~\eqref{trans} appropriately. Further, we require
$\L_{(i)}$ to be such that the ${\rm SO}(3)$ rotation $\cO_{(i)}$
represents the right-handed vector representation of ${\rm
SO}(4)$, that is $\cO_{(i)}=O_+(\L_{(i)})$. The left- and
right-handed vector representations of ${\rm SO}(4)$ have been
explicitly defined in Eq.~\eqref{Rpm}. With this choice,
$\L_{(i)}$ and $\cO_{(i)}$ together define an embedding of ${\rm
SO}(4)$ into $G_2$ as explained in Appendix~\ref{G2structures}.
This means that both rotations drop out of $\vf$ and at
$u>u_1$ we obtain
\begin{equation}
 \vf =  \sum_{a}w^a(\bar{\bzeta}_{(\t )})\wedge d\bar{\xi}_{(\t )}^a
        -d\bar{\xi}_{(\t )}^1\wedge d\bar{\xi}_{(\t )}^2
       \wedge d\bar{\xi}_{(\t )}^3
\end{equation}
Using Eq.~\eqref{phiw}, and after replacing the bared coordinates with
their rescaled counterparts, this indeed matches the bulk $G_2$
structure~\eqref{phiB} for the three coordinate identifications~\eqref{xa},
\eqref{xb}, \eqref{xc}.

\vspace{0.7cm}

Let us summarise by listing all parameters on which the above
$G_2$ structure depends. First, we have the seven radii $R^A$ of
the torus. Note that due to its dependence on the bared, rescaled
coordinates the $G_2$ structure~\eqref{phii} on the blow-ups is
also a function of these radii. In addition, for each fixed point
$(\t ,n)$, we have a radius $\r_{(\t ,n)}$, which measures the size
of the blow-up, and a ${\rm SO}(3)$ rotation $\cO_{(\t ,n)}$. As we
will see, the $G_2$ structure only depends on the normal vector
\begin{equation}
 n_{(\t, n)}^a=\cO_{(\t ,n)1a}\label{n}
\end{equation}
and, hence, only on two of the three
angles of this rotation. This normal vector parametrises the orientation
of the blow-up with respect to the bulk. The $G_2$ structure, therefore,
depends on
\begin{equation}
 R^A\; ,\qquad \r_{(\t ,n)}\; ,\qquad n_{(\t ,n)}^a\; ,
\end{equation}
where $|{\bf n}_{(\t ,n)}|=1$, which makes a total of $43$ independent
parameters, as expected.

\vspace{0.7cm}

We will now compute the family of metrics $g$, or rather vielbeins,
associated with the above $G_2$ structures using the relations~\eqref{gammadef}
and \eqref{gdef}.

For the bulk $B$, the vielbein is computed from the bulk $G_2$
structure~\eqref{phiB} and is, of course, given by
\begin{equation}
 e_A^\Au = R^A\d_A^\Au\; .
\end{equation}
Likewise, the vielbein on the blow-up $F_{(i)}$ is obtained from
the $G_2$ structure~\eqref{phii}. The essence of this calculation
has been described in the part of Appendix~\ref{G2structures} leading up to
the metric~\eqref{gass}. Here, we need to slightly generalise this
calculation to include the radii $R^A$ and the rotation of the
blow-up. To deal with this we introduce the matrices
\begin{equation}
 A_{(i)}={\rm diag}(R_{(\t )}^\m )\; ,\qquad
 B_{(i)}={\rm diag}(R_{(\t )}^a) \; .
\end{equation}
The vielbein is then given by
\begin{equation}
 \left( e_\m^\muu\right) = D(u,\r_{(i)})P(\bar{\bf z}_{(i)})A_{(i)}\; ,\qquad
 \left( e_a^\au\right) = E(u,\r_{(i)})\cO_{(i)}B_{(i)}
\end{equation}
where
\begin{eqnarray}
 D(u,\r_{(i)}) &=& \cG (u,\r_{(i)})^{-1/6}{\rm diag}\left(\sqrt{
                   \frac{\cF '(u,\r_{(i)})}{2u}},
                 \sqrt{\frac{\cF '(u,\r_{(i)})}{2u}},\frac{\sqrt{\cF (u,\r_{(i)})}}{u},
                 \frac{\sqrt{\cF (u,\r_{(i)})}}{u}\right) \\
 E(u,\r_{(i)}) &=& {\rm diag}\left( \cG (u,\r_{(i)})^{1/3},\cG (u,\r_{(i)})
                   ^{-1/6},\cG (u,\r_{(i)})^{-1/6}\right)\; .
\end{eqnarray}
Here, $\cF$ is the function defined in Eq.~\eqref{Fdef}, $\cG$
is given by
\begin{equation}
 \cG (u,\r ) \equiv \frac{\cF (u,\r)\cF '(u,\r )}{2u^3} =
            1+\frac{\r^4}{4u^3}\frac{d}{du}\left( \e^2-u\e\e '\right)+\cO (\r^8)\;
\end{equation}
and $u=|\bar{\bf z}|$. Note, that the matrix $P$, defined in Eq.~\eqref{P}
is an element of ${\rm SO}(4)$.

{}From these results, one can easily obtain the measure $\sqrt{g}$
of the associated metric. One finds that
\begin{equation}
 \sqrt{g}=\prod_A R^A \label{gB}
\end{equation}
for the bulk $B$ and
\begin{equation}
 \sqrt{g}=\cG (u,\r_{(i)})^{1/3}\prod_A R^A \label{gi}
\end{equation}
for the blow-up $F_{(i)}$.


\section{Periods, volume and K\"ahler potential}

An important ingredient in the construction of compact $G_2$
manifolds given in Ref.~\cite{joyce1} is the proof that, for sufficiently
small blow-up radii $\r_{(i)}$, the $G_2$ structure $\vf$
given in the previous section differs from its torsion-free counterpart
$\tilde{\vf}$ by an exact form $d\eta$, that is,
\begin{equation}
 \tilde{\vf}=\vf +d\eta\; . \label{phit}
\end{equation}
This implies that the periods~\eqref{periods} can, in fact, be
computed exactly from~\footnote{Here, we include a factor of $1/8$
relative to Eq.~\eqref{periods} to take care of the eight-fold
over-counting in the upstairs picture.}
\begin{equation}
 a^I=\frac{1}{8}\int_{C^I}\vf
\end{equation}
using the $G_2$ structure $\vf$ with small torsion. Let us now carry
this out starting with the bulk cycles $C^A$ defined in Eq.~\eqref{CA}
and the associated periods, which we denote by $a^A$. Since these
cycles are entirely contained within the bulk, we can use the bulk
expression~\eqref{phiB} for $\vf$. One easily finds
\begin{equation}
 \begin{array}{lllllllllllllll}
  a^1 &=& R_1R_2R_7 &,& a^2 &=& R_1R_3R_6 &,& a^3 &=& R_1R_4R_5 &,&
  a^4 &=& R_2R_3R_5 \\
  a^5 &=& R_2R_4R_6 &,& a^6 &=& R_3R_4R_7 &,& a^7 &=& R_5R_6R_7 &&&&\; .
 \end{array} \label{aA}
\end{equation}
Let us now move to the cycles related to the blow-ups. Recall, that
for each of the 12 fixed points $(\t ,a)$ we have three three-cycles
$C^{(\t ,n,a)}$, defined in Eq.~\eqref{Ctna}, which are localised in
the four equivalent patches $F_{(\t ,n,d)}$, where $d=1,2,3,4$. The
associated periods are denoted by $a^{(\t ,n,a)}$. With the
expression~\eqref{phii} for the $G_2$ structure on the blow-ups we
find
\begin{equation}
 a^{(\t ,n,a)} = -\frac{\p}{2}R_{(\t )}^a\r_{(\t ,n)}^2 n_{(\t ,n)}^a\; ,
 \label{ai}
\end{equation}
where we have used the definitions~\eqref{ww1}--\eqref{ww3} of the
two-forms $w^a$ and the explicit representations of the Maurer-Cartan
forms $\s^a$ in terms of angular coordinates given in Appendix~\ref{EH}.
Also note that the cycles $C^{(\t ,n,a)}$ are located at $u=0$, that is,
in a region where the four-dimensional part of the space is identical
to an Eguchi-Hanson space. Hence, we can use the exact form~\eqref{FGEH}
of the function $\cF$ associated with the Eguchi-Hanson metric. In
particular, we have $\cF (u=0,\r )=\r^2$ and $\cF '(u=0,\r )=0$.
Also, recall the definition~\eqref{n} of the normal vector
${\bf n}_{(\t ,n)}=(n_{(\t ,n)}^a)$ that parametrises the orientation
of the blow-up relative to the bulk.

\vspace{0.7cm}

Our next task is to compute the volume~\eqref{V} of the manifold $X$
with respect to the Ricci-flat metric $\tilde{g}$ associated with the
torsion-free $G_2$ structure $\tilde{\vf}$. Since we do not know
either $\tilde{g}$ or $\tilde{\vf}$ explicitly this cannot be done
exactly. However, as we will see, the volume can be computed in
a controlled approximation using the $G_2$ structure $\vf$ with
small torsion and its associated metric $g$ instead. From the
definition of $V$, Eq.~\eqref{V} and the relation~\eqref{phit}
we have
\begin{equation}
 V=\frac{1}{7}\int_X\tilde{\vf}\wedge\T (\tilde{\vf} )=\frac{1}{7}
   \int_X\vf\wedge\T (\tilde{\vf} )\; ,
\end{equation}
where the second equality follows after partial integration.
Further, one can expand the map $\T (\tilde{\vf})=\T (\vf +d\eta )$
around $\vf$ to linear order in $d\eta$. This leads to~\cite{joyce1}
\begin{equation}
 \T (\vf +d\eta )= \T (\vf )+\frac{4}{3}\star\p_1(d\eta )+
                   \star\p_7(d\eta )-\star\p_{27}(d\eta )+\cO( (d\eta )^2)
 \label{Texp}
\end{equation}
where $\p_1$, $\p_7$ and $\p_{27}$ denote the projections of
three-forms onto their components transforming as one of the
irreducible $G_2$ representations in $({\bf 7}\times {\bf 7}\times
{\bf 7}) _{\rm anti-symmetric}={\bf 1}+{\bf 7}+{\bf 27}$.
The fact~\cite{joyce1} that both $d\T (\vf )$ and the $L_2$ norm
of $d\eta$ are of order $\r_{(i)}^4$ implies that the terms in
Eq.~\eqref{Texp} linear in $d\eta$ only contribute to the volume
at order $\r_{(i)}^8$. The same is obviously the case for
quadratic and higher terms in~\eqref{Texp}. Hence, we learn that
the volume computed with the $G_2$ structure $\vf$, or,
equivalently, with its associated metric $g$ approximates the
exact volume up to terms of order $\r_{(i)}^8$, that is, we
have~\footnote{Again, we have included a factor of $1/8$ to
account for the over-counting in the upstairs picture.}
\begin{equation}
 V = \frac{1}{8}\int_X\sqrt{g}+\cO \left(\r_{(i)}^8\right)\; .
 \label{Vapp}
\end{equation}
Applying this formula to the measure~\eqref{gB}, \eqref{gi} computed
in the previous section we find
\begin{equation}
 V=\frac{1}{8}\left[\prod_AR^A-\frac{2\p^2}{3}\sum_{\t ,n}\r_{(\t ,n)}^4
    \prod_aR_{(\t )}^a\right]\; . \label{Vres}
\end{equation}
We remind the reader that $R_{(\t )}^a$ denote three of the seven radii
$R^A$, depending on the type $\t$ of the blow-up $(\t ,n)$, in analogy
with the definition of the coordinates $\bxi_{(\t )}$ in Eqs.~\eqref{xa},
\eqref{xb} and \eqref{xc}.

\vspace{0.7cm}

We are now ready to compute the K\"ahler potential. Using the
results~\eqref{aA} and \eqref{ai} for the periods, we can rewrite
the volume~\eqref{Vres} in terms of $a^A$ and $a^{(\t ,n,a)}$,
which constitute the real, bosonic parts of superfields.
We denote these superfields by $T^A$ and $U^{(\t ,n,a)}$ such that
\begin{equation}
 {\rm Re}(T^A)=a^A\; ,\qquad {\rm Re}(U^{(\t ,n,a)})=a^{(\t ,n,a)}\; .
\end{equation}
{}From Eq.~\eqref{Kdef} we then find for the K\"ahler potential
\begin{equation}
 K = -\sum_{A=1}^7\ln (T^A+\bar{T}^A)-3\ln\left[1-\frac{8}{3}
     \sum_{\t ,n,a}\frac{(U^{(\t ,n,a)}+\bar{U}^{(\t ,n,a)})^2}
     {(T^{A(\t ,a)}+\bar{T}^{A(\t ,a)})(T^{B(\t ,a)}+\bar{T}^{B(\t ,a)})}
     \right] + c\; , \label{K1}
\end{equation}
where the constant $c$ is given by
\begin{equation}
 c =6\ln (8\p)+\ln(2)\; .\label{c}
\end{equation}
The index functions $A(\t ,a),B(\t ,a)\in\{ 1,\dots ,7\}$ indicate
by which two of the seven bulk moduli $T^A$ the blow-up moduli
$U^{(\t ,n,a)}$ are divided in the K\"ahler potential~\eqref{K1}.
Their values depend on the type $\t$ and the orientation index $a$
but not on $n$. The nine possible values of these index functions
are summarised in Table 1.
\begin{table}
 \begin{center}
 \begin{tabular}{|l|l|l|l|}
  \hline
  $(\t ,a)$&$1$&$2$&$3$\\\hline
  $\a$&$(1,6)$&$(2,5)$&$(3,4)$\\\hline
  $\b$&$(1,7)$&$(3,5)$&$(2,4)$\\\hline
  $\g$&$(1,4)$&$(3,6)$&$(2,7)$\\\hline
 \end{tabular}
 \caption{Values of the index functions $(A(\t ,a),B(\t ,a))$
          specifying the bulk moduli $T^A$ by which the
          blow-up moduli $U^{(\t ,n,a)}$ are divided in the
          K\"ahler potential.}
 \end{center}
\end{table}
With the blow-up moduli $U^{(\t ,n,a)}$ being of order $\r_{(\t
,n)}^2$ (see Eq.~\eqref{ai}), and possible corrections to the volume
of order $\r_{(\t ,n)}^8$, we conclude that the leading corrections to
the K\"ahler potential~\eqref{K1} arise at order $(U/T)^4$. Hence,
Eq.~\eqref{K1} represents a viable approximation to the K\"ahler
potential if, firstly, all moduli are larger than one (in units where
the Planck length is set to one) so that the supergravity
approximation is valid and, secondly, if all blow-up moduli $U^{(\t
,n,a)}$ are small compared to $T^A$ so that corrections of order
$(U/T)^4$ can be neglected.


\section{K\"ahler metric and harmonic forms}

In this section we shall verify our result for the K\"ahler potential
(\ref{K1}) using Eqs.~\eqref{KIJ} and \eqref{KI}. These equations
relate derivatives of the K\"ahler potential to certain integrals involving
the harmonic forms $\{\Phi_I\}$. Specifically, we will focus on
the first derivative
\begin{eqnarray}
 K_{(\t ,n,a)} &=&16\frac{U^{(\t ,n,a)}+\bar{U}^{(\t ,n,a)}}
                  {(T^{A(\t ,a)}+\bar{T}^{A(\t ,a)})
                   (T^{B(\t ,a)}+\bar{T}^{B(\t ,a)})}
 \label{Ki}
\end{eqnarray}
and the component
\begin{eqnarray}
K_{(\t ,n,a)\overline{(\t ,n,a)}} &=& \frac{16}{(T^{A(\t
,a)}+\bar{T}^{A(\t ,a)})
                                  (T^{B(\t ,a)}+\bar{T}^{B(\t ,a)})}
 \label{Kii}
\end{eqnarray}
of the K\"ahler metric, here produced by differentiating the K\"ahler
potential~\eqref{K1}. We would now like to construct some of the harmonic
forms, compute the integrals in Eqs.~\eqref{KIJ} and \eqref{KI} and
verify that the results indeed coincide with the derivatives of the
K\"ahler potential given above.

\vspace{0.7cm}

Recall, that the integral basis of three-cycles we have used
consists of $\{ C^I\}=\{ C^A,C^{(\t ,n,a)}\}$, where $C^A$ are the
seven bulk cycles and $C^{(\t ,n,a)}$ represent three three-cycles for
each of the $12$ blow-ups $(\t ,n)$. We denote the dual basis of
harmonic three-forms, satisfying Eq.~\eqref{dual}, by
$\{\Phi_I\}=\{\Phi_A,\Phi_{(\t ,n,a)}\}$. Of course, we will not
be able to determine these harmonic forms exactly because we do
not know the exact Ricci-flat metric on $X$. However, we will find
suitable approximations for the localised forms $\Phi_{(\t ,n,a)}$
that allow us to compute some of the relevant integrals to the
accuracy required.

{}From Appendix \eqref{SEH} we know that smoothed Eguchi-Hanson
spaces have an anti-selfdual two-form
\begin{equation}
 \nu (\bar{\bf z}_{(i)},\r_{(i)}) =\frac{\cF '}{2\cF^2}du\wedge\s_1
                                   +\frac{1}{\cF}\s_2\wedge\s_3\; ,
\end{equation}
with the function $\cF$ as in Eq.~\eqref{Fdef}. For $u<u_0$, the
smoothing function $\e$ is one and this two-form is identical to
the known localised harmonic two-form on the Eguchi-Hanson
space~\cite{Cvetic:2000mh}. In this range the three-forms
\begin{equation}
 \Phi_{(\t ,n,a)}\simeq -\frac{2\r_{(\t ,n)}^2}{\p}\nu(\bar{\bf z}_{(\t ,n,d)}
                  ,\r_{(i)})\wedge d\x^a \label{harm}
\end{equation}
on $F_{(\t ,n,d)}$ are harmonic with respect to the metric $g$ associated
to the small-torsion $G_2$ structure $\vf$. The prefactor is chosen so that
\begin{equation}
 \frac{1}{8}\int_{C^{(\t ,n,a)}}\Phi_{(\t ,n,a)}=1\; ,
\end{equation}
as required for the dual basis. The forms~\eqref{harm} fall off as
$1/u^4$ for large $u$ but do not vanish exactly in the bulk.
Consequently, the above form is not valid in the bulk as this
would lead to non-vanishing integrals over the bulk cycles, contradicting Eq.~\eqref{dual}. We do not know how to smoothly
interpolate between the above expressions for the harmonic forms
on the blow-ups and bulk expressions that lead to vanishing
integrals over the bulk cycles. However, for some of the integrals
we need to perform, the behaviour at large $u$ will be irrelevant
and it is for these that~\eqref{harm} is useful. On the other
hand, we can find three-forms
\begin{equation}
 \vf_{(\t ,n,a)}=\frac{2}{\p}\bar{\nu}(\bar{\bf z}_{(\t ,n,d)})\wedge d\x^a
 \label{nonharm}
\end{equation}
on $F_{(\t ,n,d)}$, where
\begin{equation}
 \bar{\nu}(\bar{\bf z}) = \frac{\e '}{2}du\wedge\s_1+\e\s_2\wedge\s_3 \; ,
 \label{nb}
\end{equation}
that vanish in the bulk and are closed but not harmonic. They are correctly
normalised, that is
\begin{equation}
 \frac{1}{8}\int_{C^{(\t ,n,a)}}\vf_{(\t ,n,a)}=1
\end{equation}
and indeed satisfy
\begin{equation}
 \int_{C^A}\vf_{(\t ,n,a)}=0\; ,
\end{equation}
as they vanish identically in the bulk. As a result, they are
non-harmonic representatives of the cohomology classes specified
by $\Phi_{(\t ,n,a)}$.

\vspace{0.7cm}

Let us now compute some of the integrals in Eq.~\eqref{KIJ} and \eqref{KI}.
We start with the one we expect to reproduce the component~\eqref{Kii}
of the K\"ahler metric. Using
\begin{equation}
 \star\Phi_{(\t ,n,a)}\simeq\frac{\r_{\t ,a}^2V}{\p a^{A(\t ,a)}a^{B(\t ,a)}}
                            \e_{abc}\nu (\bar{\bf z}_{(\t ,n,d)})\wedge
                            d\x^b\wedge d\x^c
\end{equation}
on $F_{(\t ,n,d)}$ for $u<u_0$ we obtain
\begin{equation}
 \frac{1}{4V}\int_X\Phi_{(\t ,n,a)}\wedge\star\Phi_{(\t ,n,a)}
 = \frac{16}{(T^{A(\t ,a)}+\bar{T}^{A(\t ,a)})
   (T^{B(\t ,a)}+\bar{T}^{B(\t ,a)})}\; .
\end{equation}
This indeed matches the component~\eqref{Kii} of the K\"ahler metric
that we have obtained from the K\"ahler potential~\eqref{K1} exactly.
Note that the main contribution to the above integral comes from small values
of $u$. The contribution at large $u=u_0$ behaves like $\r_{(i)}^4/u_0^4$
and so can be neglected to the order in $\r_{(i)}$ we are working.
The range of small $u$ values is precisely the one where we can
trust the expression for the three-form~\eqref{harm}, which is why its use
in the present context is justified.

\vspace{0.7cm}

Next, we would like to consider the integrals
\begin{equation}
 \frac{1}{2V}\int_X\Phi_{(\t ,n,a)}\wedge\star\vf\;,
\end{equation}
which, from Eq.~\eqref{KI}, should reproduce the first derivatives
$K_{(\t ,n,a)}$ of the K\"ahler potential. Since the $G_2$
structure $\vf$ is non-vanishing in the bulk we cannot use the
expression~\eqref{harm} for $\Phi_{(\t ,n,a)}$, which is valid for
small $u$ only. However, since $d\star\vf\simeq 0$ to a good
approximation, we can evaluate the above integral with the
non-harmonic representatives $\vf_{(\t ,n,a)}$,
Eq.~\eqref{nonharm}, which are exactly zero in the bulk. Using the
expression~\eqref{phii} for $\vf$, some of the properties of the
Maurer-Cartan forms $\s^a$ listed in Appendix~\ref{EH} and the
results~\eqref{aA} and \eqref{ai} for the periods, one finds
\begin{equation}
 \frac{1}{2V}\int_X\vf_{(\t ,n,a)}\wedge\star\vf
  = 16\frac{U^{(\t ,n,a)}+\bar{U}^{(\t ,n,a)}}
                  {(T^{A(\t ,a)}+\bar{T}^{A(\t ,a)})
                   (T^{B(\t ,a)}+\bar{T}^{B(\t ,a)})}\;,
\end{equation}
which exactly reproduces the first derivative~\eqref{Ki} of the
K\"ahler potential. Hence, we have checked our main result~\eqref{K1}
for the K\"ahler potential by reproducing some of its derivatives
directly from integrals over harmonic forms.


\section{Gauge-kinetic functions and singularities}

The manifold we are considering has $b^2(X)=12$ and, hence, there
exist $12$ Abelian gauge multiplets $A^{(\t ,n)}$, one for each
blow-up $(\t ,n)$. We would now like to compute the gauge-kinetic
functions~\eqref{fJK} for these vector multiplets. First we need
explicit expressions for the basis $\{\o_{(\t ,n)}\}$ of two-forms dual
to the basis of two-cycles $\{D^{(\t ,n)}\}$ defined in
Eq.~\eqref{DI}. Since the integrals~\eqref{cIJK} are topological we
can use any representatives for these dual two-forms, including non-harmonic
ones. A convenient choice is provided by
\begin{equation}
 \o_{(\t ,n)} = \frac{2}{\p}\bar{\n}(\bar{\bf z}_{(\t ,n,d)})
\end{equation}
on $F_{(\t ,n,d)}$, where $\bar{\n}$ is defined in Eq.~\eqref{nb}.  It
is immediately clear from this form that the gauge kinetic function
$f_{(\t ,n)(\t ',n')}$ coupling the field strengths $F^{(\t ,n)}$ and
$F^{(\t ',n')}$ is non-zero only for gauge fields of the same type,
that is when $\t =\t '$ and $n=n'$. Moreover, their value only depends
on the type $\t$ but not on $n$, so that we have three types of gauge
couplings $f_{(\t )}$, where $\t = \a ,\b ,\g$. A short calculation
using Eqs.~\eqref{cIJK} and \eqref{fJK} shows that
\begin{equation}\label{gaugekin}
 f_{(\t )}=\left\{\begin{array}{lll}
   T^7&{\rm for}&\t =\a\\
   T^6&{\rm for}&\t =\b\\
   T^5&{\rm for}&\t =\g
 \end{array}\right.
\end{equation}
Note that $c_{IJK}$ is only non-zero when the modulus $T^A$ corresponds to
a bulk harmonic form $\Phi_A$ living on the three torus parallel to the
blow-up under consideration. That is, $A=7$ for $\t=\a$;  $A=6$ for
$\t=\b$ and $A=5$ for $\t=\g$, which is precisely the pattern emerging
in \eqref{gaugekin}.

\vspace{0.7cm}

We now move on to discuss the four-dimensional effective theory when
some of the blow-ups in our $G_2$ manifold $X$ collapse back to an
orbifold singularity.

Generally, a co-dimensional four singularity that is locally of the form
$\CM^2/\ZM_n\times Y$ with a three-cycle $Y\subset X$, that is,
a singularity of type $A_{n-1}$, leads to additional massless gauge fields
with gauge group ${\rm SU}(n)$. These gauge fields are localised on the
seven-dimensional manifold $Y\times M_4$, where $M_4$ is
four-dimensional space-time. They can be incorporated by adding to
11-dimensional supergravity on $X\times M_4$ the action
\begin{equation}
 S_{\rm YM} =- \frac{1}{4\l^2}\int_{Y\times M_4}\left[
              d^7x\sqrt{-g_7}\,{\rm tr}F^2+C\wedge{\rm tr}(F\wedge F)
              \right] \label{YM}
\end{equation}
for the ${\rm SU}(n)$ gauge field $A$ with associated field strength
$F$. Here $g_7$ is the induced metric on $Y\times M_4$. In
Ref.~\cite{Friedmann:2002ty}, the gauge coupling $\l$ has been
determined as
\begin{equation}
 \l = (16\p^2)^{2/3}\; ,
\end{equation}
in units where the 11-dimensional Newton constant $\k$ is set to
one. The second term in this action has been inferred from anomaly
cancellation arguments in Ref.~\cite{Witten:2001uq}. This term allows
one to explicitly determine the four-dimensional gauge-kinetic function
$f$. Let us expand the three-cycle $Y$ in terms of our basis $\{
C^I\}$ of three-cycles as
\begin{equation}
 Y=\sum_Im_IC^I\label{Y}
\end{equation}
with integer expansion coefficients $m_I$. Then reducing the
action~\eqref{YM} by integrating over $Y$ and using the
expansion~\eqref{C} for $C$ one finds
\begin{equation}
 {\rm Im}(f)=\sum_Im_I\n^I\; ,\qquad  {\rm Re}(f)={\rm vol}(Y)\; .
\end{equation}
We recall that the axions $\n^I$ are the imaginary parts
of the chiral superfields $T^I$. Holomorphicity of the gauge-kinetic
function then tells us that
\begin{equation}
 f=\sum_Im_IT^I\; .\label{f}
\end{equation}
By combining the expression for ${\rm Re}(f)$ with the
one deduced from holomorphicity, Eq.~\eqref{f}, we learn that
the volume of $Y$ can be written as
\begin{equation}
 {\rm vol}(Y)=\sum_I m_I a^I \label{Ya}
\end{equation}
in terms of the metric moduli $a^I$.

\vspace{0.7cm}

We would now like to apply these general remarks to the specific
$G_2$ manifold $X$ we have focused on in this paper. Each of the
$12$ blow-ups of this manifold originates from an $A_1$
singularity, so a collapse of each of these blow-ups will lead
to an ${\rm SU}(2)$ gauge field in the four-dimensional effective
theory. What does the moduli K\"ahler potential look like when
some of the blow-ups $(\t ,a)$ have collapsed? The natural
conjecture is that the K\"ahler potential is still of the
form~\eqref{K1} but with the terms corresponding to collapsed
blow-ups set to zero. Formally, this can be achieved by the
blow-down operation ${\rm Re}(U^{\t ,n,a})\rightarrow 0$ for all
singular $(\t ,a)$ in Eq.~\eqref{K1}.

Let us next determine the gauge-kinetic functions for these
${\rm SU}(2)$ gauge fields. Their precise form depends on the type $\t
=\a ,\b ,\g$ of the orbifold singularity. Comparing the definition
of the orbifold action in Eqs.~\eqref{a}, \eqref{b} and \eqref{g}
with the definition~\eqref{CA} of the basis $\{ C^I\}$ of three
cycles one finds that
\begin{equation}
 Y=\left\{\begin{array}{lll}
   C^7&{\rm for}&\t =\a\\
   C^6&{\rm for}&\t =\b\\
   C^5&{\rm for}&\t =\g
 \end{array}\right.
\end{equation}
The gauge-kinetic function is then obtained from Eqs.~\eqref{Y} and
\eqref{f} and it coincides with the result~\eqref{gaugekin} that we
found for the Abelian gauge multiplets in the smooth case. This is not
unexpected given that the Abelian vector multiplet present in the
smooth case corresponds to the ${\rm U}(1)$ vector multiplet contained
within the ${\rm SU}(2)$ that forms in the singular limit.


\vspace{1cm}

\noindent
{\Large\bf Acknowledgements}\\ We would like to thank Philip Candelas,
Jan Louis, Paul Saffin and Kelly Stelle for helpful discussion. A.~L.~is
supported by a PPARC Advanced Fellowship, and S.~M.~by a PPARC
Postgraduate Studentship.\\


\vskip 1cm
\appendix{\noindent\Large \bf Appendix}
\renewcommand{\theequation}{\Alph{section}.\arabic{equation}}
\setcounter{equation}{0}


\section{Blow-up and Eguchi-Hanson metric} \label{EH}

In this Appendix, we collect some standard material on the blow-up
of $\CM^2/\ZM^2$, and the associated Eguchi-Hanson metric, that will
be used in our calculations. We mainly follow
Refs.~\cite{gh}-\cite{lebrun}.

\vspace{0.7cm}

Let us first recall how to construct the blow-up of the origin in
$\CM^2/\ZM_2$, where the $\ZM_2$ action on the complex coordinates
$Z=(z_1,z_2)$ is defined by $(z_1,z_2)\rightarrow (-z_1,-z_2)$.
More precisely, we focus on a four-dimensional ball $B_\s$, with
radius $\s$, centred around the origin of $\CM^2$. Introducing
homogeneous coordinates $L=[l_1,l_2]$ on $\CM\PM^1$ the blow-up
$U\subset B^4_\s\times \CM\PM^1$ can be defined as
\begin{equation}
 U = \{(Z,L)\in B_\s\times\CM\PM^1 | z_1l_2=z_2l_1\}\; . \label{blow-up}
\end{equation}
If this definition is extended to all of $\CM^2$, the resulting
space can be identified with $T^*\CM\PM^1$, the cotangent bundle
over $\CM\PM^1$. The blow-down projection $\p$ is defined by $\p
(Z,l)=Z$. For $z_1\neq 0$ or $z_2\neq 0$, $\p^{-1}(z)$ consists of
a single point, as the condition in~\eqref{blow-up} shows. Hence,
away from the origin the blow-up looks like $\CM^2$ locally. At
the origin, $z_1=z_2=0$, on the other hand, we have
\begin{equation}
 E\equiv\p^{-1}(0)\simeq \CM\PM^1\; .
\end{equation}
The cycle $E$ is called the exceptional divisor of the blow-up. The manifold
$U$ can be covered by two coordinate charts, $U_1=\{ l_1\neq 0\}$ and
$U_2=\{ l_2\neq 0\}$, with associated coordinates
\begin{equation}
 \frac{l_2}{l_1}=\frac{z_2}{z_1}\; ,\qquad z_1 \label{c1}
\end{equation}
for $U_1$ and
\begin{equation}
\frac{l_1}{l_2}=\frac{z_1}{z_2}\; ,\qquad z_2 \label{c2}
\end{equation}
for $U_2$.

\vspace{0.4cm}

Subsequently, we will need both real and polar coordinates on $\CM^2$.
We introduce real coordinates ${\bf z}=(z^\m)=(x,y,z,t)$, where
$\m ,\n ,\dots = 0,1,2,3$, by
\begin{equation}
 z_1=x+iy\; ,\qquad z_2=z+it
\end{equation}
and polar coordinates $(u,\q,\phi,\psi )$ by
\begin{equation}
 z_1 = u\cos\frac{\q}{2}\exp\left(\frac{i}{2}(\psi +\phi )\right)\; ,\qquad
 z_2 = u\sin\frac{\q}{2}\exp\left(\frac{i}{2}(\psi -\phi )\right)\; .
\end{equation}
Here, the three angular coordinates vary within the ranges
\begin{equation}
 \q\in [0,\p ]\; ,\qquad \phi\in [0,2\p ]\; ,\qquad \psi\in [0,4\p ]\; .
  \label{range}
\end{equation}
The action of $\ZM_2$ in polar coordinates is given by
$\psi\rightarrow \psi +2\p$ with $u$, $\q$ and $\phi$ unchanged.
In a ``downstairs" picture, where one works with the fundamental
domain only, the range of $\psi$ should be restricted to $\psi\in
[0,2\p ]$. Here, we will usually use the ``upstairs" picture and,
consequently, work with the full range as in Eq.~\eqref{range}.
Note that, from Eqs.~\eqref{c1} and \eqref{c2}, the angles $\q$
and $\phi$ can be considered good coordinates on the exceptional
divisor $E$ so that we can write, in polar coordinates
\begin{equation}
 E = \{ u=0,\q\in [0,\p ],\phi\in [0,2\p ]\}\; .
\end{equation}
This interpretation of the exceptional divisor will be particularly
useful for explicit calculations.

\vspace{0.4cm}

To proceed further, we need to recall some ${\rm SO}(4)$ group properties.
We first introduce a basis of the ${\rm SO}(4)$ Lie algebra consisting
of left- and right-handed generators $T_\pm^\au$. They take the explicit form
\begin{equation}
 (T_\pm^\au)_{0\bu}=\d^\au_\bu\; ,\qquad
 (T_+^\au)_{\bu\cu}=-(T_-^\au)_{\bu\cu}={\e^\au}_{\bu\cu}\; ,
 \label{Tdef}
\end{equation}
where $\au ,\bu ,\dots = 1,2,3$, and they satisfy the standard
commutation relations $[T_\pm^\au ,T_\pm^\bu ]=2{\e^{\au\bu}}_\cu
T_\pm^\cu$ and $[T_+^\au ,T_-^\bu ]=0$. It is useful to add the unit
matrix as $T_\pm^0$ to obtain a `` covariant" version
\begin{equation}
 (T_\pm^\muu ) = ({\bf 1}_4,T_\pm^\au )\; ,
\end{equation}
where $\muu ,\nuu ,\dots = 0,1,2,3$. We also note that the left- and
right-handed vector representations $R_\pm$ of ${\rm SO}(4)$ are
obtained from the relations
\begin{equation}
 \L^TT_\pm^\muu\L = {R_\pm (\L )^\muu}_\nuu T_\pm^\nuu \label{Rpm}
\end{equation}
where $\L\in{\rm SO}(4)$ and
\begin{equation}
 R_\pm (\L )=\left(\begin{array}{cc}1&0\\0&O_\pm (\L )\end{array}\right)
\end{equation}
with $O_\pm (\L )\in{\rm SO}(3)$.

\vspace{0.7cm}

We can now introduce the Maurer-Cartan one-forms $\s^\au$ on
$S^3\simeq {\rm SU}(2)$ satisfying
\begin{equation}
 d\s^\au = {\e^\au}_{\bu\cu}\s^\bu\wedge\s^\cu\; . \label{MC}
\end{equation}
In terms of the one-forms $(\e^\muu )=(du,u\s^\au )$ on $\CM^2$,
they can be explicitly defined as
\begin{equation}
 \e^\muu = \frac{1}{|{\bf z}|}{\bf z}^TT_+^\muu d{\bf z}\; .\label{eps+}
\end{equation}
Alternatively, in terms of the left-handed generators $T_-^\au$
we can write
\begin{equation}
 \e^\muu = {P({\bf z})^\muu}_\n dz^\n\; , \label{eps-}
\end{equation}
where the matrix $P({\rm z})$ takes the form
\begin{equation}
 P({\bf z}) = \frac{1}{|{\bf z}|}\left(\begin{array}{rrrr}
              x&y&z&t\\-y&x&-t&z\\-z&t&x&-y\\-t&-z&y&x
              \end{array}\right)
            = \frac{1}{|{\bf z}|}\sum_\m z^\m T_-^\m\; . \label{P}
\end{equation}
We remark that $P({\bf z})\in {\rm SO}(4)$. One can verify by
straightforward calculation that the so-defined forms $\s^\au$ indeed
satisfy the ${\rm SO}(3)$ Maurer-Cartan equation~\eqref{MC}. Further,
it is easy to see from Eqs.~\eqref{eps+} and \eqref{Rpm} that they
transform in the right-handed vector representation of
${\rm SO}(4)$, that is,
\begin{equation}
 \s^\au (\L{\bf z}) = {{O_+(\L )}^\au}_\bu\;\s^\bu ({\bf z})\; .
\end{equation}

In polar coordinates, the forms $\s^\au$ can be written
as~\footnote{We will write explicit indices $\au$ on $\s^\au$ as
lower indices for notational convenience.}
\begin{eqnarray}
 \s_1 &=& \frac{1}{2}(\cos\q d\phi + d\psi ) \nn \\
 \s_2 &=& \frac{1}{2}(\cos\psi d\q +\sin\q\sin\psi d\phi ) \\
 \s_3 &=& \frac{1}{2}(\sin\psi d\q - \sin\q\cos\psi d\phi )\nn \; .
\end{eqnarray}
We also collect the wedge products
\begin{eqnarray}
 \s_2\wedge\s_3 &=& -\frac{1}{4}\sin\q d\q\wedge d\phi \nn \\
 \s_3\wedge\s_1 &=& \frac{1}{4}(\cos\q\sin\psi d\q\wedge d\phi+\sin\psi
                    d\q\wedge d\psi -\sin\q\cos\psi d\phi\wedge d\psi ) \\
 \s_1\wedge\s_2 &=& -\frac{1}{4}(\cos\q\cos\psi d\q\wedge d\phi +
                    \cos\psi d\q\wedge d\psi + \sin\q\sin\psi d\phi\wedge
                    d\psi )\nn
\end{eqnarray}
Note that the first of these forms, $\s_2\wedge\s_3$ is
well-defined on the exceptional divisor $E$ because it depends on
$\q$ and $\phi$ only. Therefore it can be extended to the blow-up
$U$. It is, in fact, proportional to the volume form on $E\simeq
S^2$. The other two wedge products, and indeed $\s_2$ and $\s_3$
themselves, do depend on $\psi$ and are, therefore, not
well-defined at the origin. Nevertheless, as we will see below,
they can appear in certain forms that are well-defined on the
blow-up $U$ as long as they are multiplied with a function of the
radial coordinate $u$ that vanishes at the origin $u=0$. Further,
we have
\begin{equation}
 \s_1\wedge\s_2\wedge\s_3 = -\frac{1}{8}\sin\q d\q\wedge d\phi\wedge d\psi
\end{equation}
and
\begin{equation}
 \s_2^2+\s_3^2 = \frac{1}{4}(d\q^2+\sin^2\q d\phi^2)\; . \label{gS2}
\end{equation}
The flat volume form on $\CM^2$ can then be written as
\begin{equation}
 dx\wedge dy\wedge dz\wedge dt = -\frac{u^3}{8}\sin\q du\wedge d\q\wedge d\phi
                                 \wedge d\psi
                               = u^3du\wedge \s_1\wedge\s^2\wedge\s_3\; .
 \label{d4x}
\end{equation}

\vspace{0.7cm}

Let us briefly recall the definition of a hyperk\"ahler space before
proceeding with our example. A hyperk\"ahler space is a $4m$--dimensional
Riemannian manifold with metric $g$, called the hyperk\"ahler metric, and
a triplet $J^\au$ of covariantly constant complex structures satisfying the algebra
\begin{equation}
 J^\au J^\bu = -{\bf 1}\d^{\au\bu}+{\e^{\au\bu}}_\cu J^\cu\; . \label{quat}
\end{equation}
As a consequence, the K\"ahler forms $w^\au$ associated with $J^\au$ via
\begin{equation}
 w^\au_{\m\n} = {(J^\au )_\m}^\r g_{\r\n}
\end{equation}
are also covariantly constant and, hence, closed and co-closed, that is $dw^\au =0$
and $d\star w^\au =0$.

\vspace{0.7cm}

The Eguchi-Hanson hyperk\"ahler metric on the blow-up $U$ can be obtained
from the K\"ahler potential
\begin{equation}
 \cK = \sqrt{u^4+\r^4}+2\r^2\ln u -\r^2\ln\left(\sqrt{u^4+\r^4}+\r^2\right)\; ,
 \label{KEH}
\end{equation}
where
\begin{equation}
 u^2=|z_1|^2+|z_2|^2\;
\end{equation}
and $\r$ is a real parameter that measures the radius of the
exceptional divisor $E$. For such ${\rm U}(2)$ invariant K\"ahler
potentials, which depend on the complex coordinates through the
radial coordinate only, it is useful to introduce two auxiliary
functions
\begin{equation}
 \cF= \frac{u\cK '}{2}\; ,\qquad \cG = \frac{\cF \cF'}{2u^3}\; ,\label{FGdef}
\end{equation}
where the prime denotes the derivative with respect to $u$.
For the concrete example~\eqref{KEH} we find for these functions
\begin{equation}
 \cF = \sqrt{u^4+\r^4}\; ,\qquad \cG = 1\; . \label{FGEH}
\end{equation}
By straightforward calculation, the metric associated with the K\"ahler
potential~\eqref{KEH} can be obtained as
\begin{equation}
 ds_{\rm EH}^2 = \frac{\cF '}{2u}du^2+\frac{u\cF '}{2}\s_1^2+\cF (\s_2^2+\s_3^2)\; .
 \label{gEH}
\end{equation}
where $\cF$ is given in Eq.~\eqref{FGEH}. This metric is Ricci-flat and
constitutes a hyperk\"ahler metric on the blow-up $U$, that is, it has
associated with it a triplet of three integrable complex structures and
three covariantly constant K\"ahler potentials. We will present these
K\"ahler forms explicitly below.

The metric~\eqref{gEH} takes a more familiar form when written in the radial
coordinate $r$ defined by
\begin{equation}
 r^2 = \cF = \sqrt{u^4+\r^4}\; .
\end{equation}
It then turns into
\begin{equation}
 ds_{\rm EH}^2 = \frac{dr^2}{1-\frac{\r^4}{r^4}}+r^2\left[\left(1-
                 \frac{\r^4}{r^4}\right)\s_1^2+\s_2^2+\s_3^2\right]\; .
\end{equation}
The coordinate $r$ is restricted to $r\in [\r ,\infty ]$ and, hence, its
range depends on the parameter $\r$. Throughout this paper, we will work with
the coordinate $u$ whose range $u\in [0,\infty ]$ is independent of $\r$.

On the exceptional divisor, at $u=0$, the metric~\eqref{gEH} degenerates into
\begin{equation}
 ds^2_{\rm EH}(u=0) = \r^2 (\s_2^2+\s_3^2) = \frac{\r^2}{4}(d\q^2+\sin^2\q
                      d\phi^2)\; ,
\end{equation}
where the second equality follows from Eq.~\eqref{gS2}. This is indeed
the standard metric on a two-sphere $E\simeq S^2$ with radius $\r /2$.

The vierbein one-forms $e^\muu$ associated with the metric~\eqref{gEH}
are given by
\begin{equation}
 (e^\muu )=\left(\sqrt{\frac{\cF '}{2u}}du,\sqrt{\frac{u\cF '}{2}}\s_1,
           \sqrt{\cF}\s_2,\sqrt{\cF}\s_3\right)\; . \label{vEH}
\end{equation}

\vspace{0.7cm}

The K\"ahler form $w_1$, associated with the K\"ahler potential~\eqref{KEH},
takes the form
\begin{equation}
 w_1 = \frac{\cF '}{2}du\wedge\s_1 + \cF\s_2\wedge\s_3\; ,
\end{equation}
where $\cF$ is as given in Eq.~\eqref{FGEH}. As discussed above,
$\s_2\wedge\s_3$ can be understood as a form on $U$, while $\cF (u=0,\r
)=0$. Therefore, $w_1$ is indeed a well-defined form on the blow-up
$U$. The two other K\"ahler forms, which must exist on a hyperk\"ahler
space, are given by
\begin{equation}
 w_2 = {\rm Re}(dz_1\wedge dz_2)\; , \qquad w_3 = {\rm Im}(dz_1\wedge dz_2)
 \label{w23}
\end{equation}
and can be expressed in polar coordinates as
\begin{eqnarray}
 w_2 &=& u\, du\wedge\s_2+u^2\s_3\wedge\s_1 \label{w2}\\
 w_3 &=& u\, du\wedge\s_3+u^2\s_1\wedge\s_2\; . \label{w3}
\end{eqnarray}
Note that these forms vanish on the exceptional divisor at $u=0$
and are, therefore, well-defined on the full blow-up $U$, although
this is not the case for some of their constituent forms, such as
$\s_2$, $\s_3$, $\s_3\wedge\s_1$ and $\s_1\wedge\s_2$. From the
relation~\eqref{MC} all three forms $w^\au$ are closed, as they
should be. Moreover, in terms of the vierbein~\eqref{vEH}, they
can be written as
\begin{equation}
 w^\au = e^0\wedge e^\au +\frac{1}{2}{\e^\au}_{\bu\cu}e^\bu\wedge e^\cu\; ,
 \label{we}
\end{equation}
which implies their self-duality, $\star w^\au = w^\au$, by virtue of the
identity
\begin{equation}
 \star\left( e^\muu\wedge e^\nuu\right) = \frac{1}{2}{e^{\muu\nuu}}_{\ru\su}
                                          e^\ru\wedge e^\su\; .
\end{equation}
Hence, the forms $w^\au$ are closed and co-closed as expected on a
hyperk\"ahler manifold. One can also explicitly verify that the
complex structures $J^\au$ associated with $w^\au$ do indeed
satisfy the algebra~\eqref{quat}.

Let us discuss the asymptotic form of these K\"ahler forms for $u\gg\r$.
As the vierbein approaches the flat space expression
$e^\m\rightarrow dz^\m$ in this limit, we have from Eq.~\eqref{we}
\begin{equation}
 w^a\rightarrow\bar{w}^a\equiv dz^0\wedge dz^a +\frac{1}{2}
   {\e^a}_{bc}dz^b dz^c = \frac{1}{2}(T_+^a)_{\m\n}dz^\m dz^\n\; .
 \label{wb}
\end{equation}
These are precisely the three constant K\"ahler forms associated
with flat space (regarded as a hyperk\"ahler space).

\vspace{0.7cm}

There also exists a closed anti-selfdual two-form~\cite{Cvetic:2000mh}
$\nu$ on $U$ which can be written as
\begin{equation}
 \nu = \frac{1}{\cF^2}(e^0\wedge e^1-e^2\wedge e^3)
     = \frac{\cF '}{2\cF^2}du\wedge \s_1 + \frac{1}{\cF}\s_2\wedge \s_3\; ,
  \label{nu}
\end{equation}
with $\cF$ as in Eq.~\eqref{FGEH}. Unlike the above K\"ahler forms
this form is ``localised", that is, it falls off as $1/u^4$ for
$u\gg\r$.


\section{Smoothed Eguchi-Hanson spaces} \label{SEH}

The Eguchi-Hanson space discussed in the previous Appendix
approaches flat space asymptotically. However, what is really
needed for the construction of $G_2$ manifolds are smoothed
versions of this space which become exactly flat for sufficiently
large radius. In this Appendix, we will discuss such smoothed
Eguchi-Hanson spaces, which interpolate between Eguchi-Hanson space
at small radius and flat space at large radius, following
Ref.~\cite{joyce1}.

 \vspace{0.7cm}

In this context, it is useful to start by analysing general K\"ahler
potentials
\begin{equation}
 \cK =\cK (u)\; ,\qquad u^2=|z_1|^2+|z_2|^2 \label{Kgen}
\end{equation}
on a four-dimensional ball $B_\s$ with radius $\s$ around the
origin of $\CM^2/\ZM_2$ (or on the blow-up $U$ of this space,
depending on the properties of $\cK$). As we will see, many of the
properties and relations we require can be obtained within this
general framework. Two examples for such K\"ahler potentials are
provided by the K\"ahler potential~\eqref{KEH} associated with the
Eguchi-Hanson space and that for flat space
\begin{equation}
 \cK = u^2\; .
\end{equation}
More complicated examples will be given below but for now we keep
$\cK = \cK (u)$ arbitrary.

As before, it is helpful to introducing the auxiliary functions
$\cF$ and $\cG$ by
\begin{equation}
 \cF = \frac{u\cK '}{2}\; ,\qquad \cG = \frac{\cF\cF'}{2u^3}\; ,\label{FGdef1}
\end{equation}
where the prime denotes the derivative with respect to $u$. The metric
associated with~\eqref{Kgen} is then given by
\begin{equation}
 ds_{\rm EH}^2 = g_{\m\n}dz^\m dz^\n
               = \frac{\cF '}{2u}du^2+\frac{u\cF '}{2}\s_1^2+\cF(\s_2^2+\s_3^2)\; ,
 \label{gEH1}
\end{equation}
and the vierbein reads
\begin{equation}
 (e^\muu )=\left(\sqrt{\frac{\cF '}{2u}}du,\sqrt{\frac{u\cF '}{2}}\s_1,
           \sqrt{\cF}\s_2,\sqrt{\cF}\s_3\right)\; . \label{vEH1}
\end{equation}
It is interesting to note that the measure derived from this
metric, given by
\begin{equation}
 \sqrt{{\rm det}(g)}=\cG=\frac{1}{4u^3}\frac{d}{du}\left( \cF^2\right)\;,
 \label{detK}
\end{equation}
only depends on the derivative of $\cF^2$. With the factor $u^3$ from
$\mathrm{d}^4z$ cancelling that in $\cG$, the volume ${\rm
vol}(u_0,u_1)$ of the part of the space defined by $u\in [u_0,u_1]$
takes the form
\begin{equation}
 {\rm vol}(u_0,u_1)=\int_{u\in [u_0,u_1]}\sqrt{g}\, d^4z
                   = \frac{\p^2}{2}\left(\cF (u_1)^2-\cF (u_0)^2\right)\; ,
 \label{vol}
\end{equation}
where we have used Eq.~\eqref{d4x}. Hence, remarkably, this volume can be
calculated from the first derivative of the K\"ahler potential directly
without the need for explicit integration.

The K\"ahler form $w_1$ derived from~\eqref{Kgen} is
\begin{equation}
 w_1 = \frac{\cF '}{2}du\wedge\s_1 + \cF\s_2\wedge\s_3=e^0\wedge e^1+e^2\wedge e^3
 \; , \label{w1}
\end{equation}
which, from Eq.~\eqref{MC} is closed, as it should be, and
selfdual. This form is well-defined on $\CM^2/\ZM^2$, and well-defined
on the blow-up $U$ as long as $\cF '(u=0,\r )=0$.  We can still
introduce a triplet $w^\au$ of closed two-forms by defining $w_2$ and
$w_3$ as in Eqs.~\eqref{w23}, \eqref{w2} and \eqref{w3}. However, the
so-defined forms are in general no longer selfdual and, hence, no
longer co-closed. This reflects the fact that we are merely working
with K\"ahler spaces and only certain K\"ahler potentials of the
form~\eqref{Kgen} correspond to hyperk\"ahler spaces.

Remarkably, the generalisation of the anti-selfdual closed form~\eqref{nu}
can be found for the general K\"ahler potential~\eqref{Kgen}. It is
given by
\begin{equation}
 \nu = \frac{1}{\cF^2}(e^0\wedge e^1-e^2\wedge e^3)
     = \frac{\cF '}{2\cF^2}du\wedge \s_1 + \frac{1}{\cF}\s_2\wedge \s_3\; ,
  \label{nu1}
\end{equation}
and is well-defined on $\CM^2/\ZM^2$ as long as the function $\cF$ is everywhere different from zero. If in addition $\cF '(u=0,\r )=0$ it is
also well-defined on the blow-up $U$.

\vspace{0.7cm}

As we have seen, all the relevant objects on a K\"ahler space defined
by a K\"ahler potential of the form~\eqref{Kgen} are determined in terms
of the two functions $\cF$ and $\cG$, defined in Eq.~\eqref{FGdef1}. Let
us now apply the above formalism to a number of examples by
computing these functions in each case.

We start with the trivial case of flat space
\begin{equation}
 \cK = u^2
\end{equation}
which, of course, constitutes a hyperk\"ahler space. One finds that
\begin{equation}
 \cF = u^2\; ,\qquad \cG = 1\; .
\end{equation}
{}From Eq.~\eqref{vol} we find for the total volume of the space
\begin{equation}
 {\rm vol}(0,\s )=\frac{\p^2}{2}\s^4\; .
\end{equation}

For the Eguchi-Hanson hyperk\"ahler space the K\"ahler potential reads
\begin{equation}
 \cK = \sqrt{u^4+\r^4}+2\r^2\ln u -\r^2\ln\left(\sqrt{u^4+\r^4}+\r^2\right)\; ,
 \label{KEH1}
\end{equation}
which leads to
\begin{equation}
 \cF = \sqrt{u^4+\r^4}\; ,\qquad \cG = 1\; . \label{FGEH1}
\end{equation}
The total volume is identical to the one for flat space, that is
\begin{equation}
 {\rm vol}(0,\s )=\frac{\p^2}{2}\s^4\; .
\end{equation}

The third example~\cite{joyce1} is defined by the K\"ahler potential
\begin{equation}
 \cK = \sqrt{u^4+\e (u)^2\r^4}+2\e (u)\r^2\ln u -\e (u)\r^2\ln\left(\sqrt{u^4+
       \e (u)^2\r^4}+\e (u)\r^2\right)\; ,
 \label{SKEH}
\end{equation}
where $\e (u)$ is a function with
\begin{equation}
 \e (u) =\left\{\begin{array}{ll}1&{\rm if}\; u\leq u_0\\
         0&{\rm if}\; u\geq u_1\end{array}\right.\; ,
\end{equation}
where $u_0$ and $u_1$ are two characteristic radii satisfying
$\r\ll u_0< u_1 <\s$.  Hence, the space described by the K\"ahler
potential~\eqref{SKEH} is identical to the Eguchi-Hanson space for
radii $u\leq u_0$ and identical to flat space for $u\geq u_1$,
that is, it interpolates between the Eguchi-Hanson space and flat
space. Although this space interpolates between two hyperk\"ahler
spaces it is not a hyperk\"ahler space by itself. Accordingly the
forms $w_2$ and $w_3$ are no longer co-closed in the ``collar''
region $u\in [u_0,u_1 ]$. For the functions $\cF$ and $\cG$ we
find, to order $\r^6$ in the blow-up radius
\begin{eqnarray}
 \cF &=& u^2+\frac{1}{2u^2}(\e^2-u\e\e ')\r^4 +\cO (\r^8) \label{SF}\\
 \cG &=& 1+\frac{\r^4}{4u^3}\frac{d}{du}\left( \e^2-u\e\e '\right)+\cO (\r^8)\; .\label{SG}
\end{eqnarray}
These functions interpolate between their counterparts for
Eguchi-Hanson and flat space, as they should. Moreover, the
correction terms arising in the collar $u\in [u_0,u_1]$ are at least of order
$\r^4$ and proportional to derivatives of the interpolating function
$\e$. Hence, this space can be thought of as being close to
hyperk\"ahler as long as the blow-up radius $\r$ is sufficiently small
compared to one and the function $\e$ is slowly varying. The function $\cG$
will eventually be the crucial ingredient in computing the volume of the
$G_2$ space. Note that the $\cO (\r^4)$ correction to
$\cG$ can be written as a total derivative. From Eq.~\eqref{vol},
the total volume is given by
\begin{equation}
 {\rm vol}(0,\s )=\frac{\p^2}{2}(\s^4 - \r^4) \label{Svol}
\end{equation}
and so is independent of the precise form of the smoothing function
$\e$. This property, which we will recover in slightly different
circumstances when we compute the volume of the $G_2$ manifold, is
crucial for our calculation. The second term in~\eqref{Svol} represents
the amount subtracted from the volume due to the presence of the blow-up
and it equals the volume of a four-dimensional ball with radius $\r$.


\section{$G_2$ structures} \label{G2structures}

This Appendix collects some useful information on the group $G_2$ and $G_2$
structures on seven-dimensional manifolds. We also describe the specific
example of a $G_2$ structure on $U\times T^3$ where $U$ is the
blow-up of $\CM^2 /\ZM_2$ described in Appendix~\ref{EH}. In parts, we
follow Ref.~\cite{joyce1,joyceb}.

\vspace{0.7cm}

We start by defining the flat $G_2$ structure
\begin{eqnarray}
 \bar{\vf} &=& dx^1\wedge dx^2\wedge dx^7 + dx^1\wedge dx^3\wedge dx^6
            + dx^1\wedge dx^4\wedge dx^5\nn \\
            && + dx^2\wedge dx^3\wedge dx^5 + dx^4\wedge dx^2\wedge dx^6
               + dx^3\wedge dx^4\wedge dx^7 + dx^5\wedge dx^6\wedge dx^7
 \label{phib}
\end{eqnarray}
on $\RM^7$ with coordinates ${\bf x} = (x^A)$, where $A,B,\dots =
1,\dots ,7$.  Consider elements of the seven-dimensional special
orthogonal group, $g\in {\rm SO}(7)$ acting linearly on ${\bf x}$. The
group $G_2$ can be defined as the subgroup consisting of elements
$g\in {\rm SO}(7)$ that leave the three-form $\bar{\vf}$ invariant
under the linear action on ${\bf x}$.

\vspace{0.7cm}

For our purpose, it is useful to split up the seven-dimensional
coordinates ${\bf x}$ into a four-dimensional part with coordinates
$\bzeta =\z^\m$, where $\m ,\n ,\dots = 0,1,2,3$, and a complementary
three-dimensional part with coordinates $\bxi =\x^a$, where $a,b,\dots
= 1,2,3$. Let us also introduce the two-forms
\begin{equation}
 \bar{w}^a = \frac{1}{2}(T_+^a)_{\m\n}d\z^\m d\z^\n
 \label{wb1}
\end{equation}
on the four-dimensional part of the space. The matrices $T_+^a$ have been defined
in Eq.~\eqref{Tdef}. We have already encountered these two-forms in Eq.~\eqref{wb}
as the triplet of K\"ahler forms on $\RM^4$. For appropriate identifications
of the seven-dimensional coordinates ${\bf x}$ with $\bzeta$ and $\bxi$
the three-form $\bar{\vf}$ can be written as
\begin{equation}
 \bar{\vf} = \sum_{a=1}^3\bar{w}^a\wedge d\xi^a - d\x^1\wedge d\x^2\wedge d\x^3\; .
 \label{phiw}
\end{equation}
A particular example of an identification for which the above
relation holds is provided by $\bzeta = (x^1,x^2,x^3,x^4)$ and
$\bxi = (x^7,x^6,x^5)$. However there are other possibilities,
obtained by suitable permutations of the coordinates, that we will
encounter in the construction of the $G_2$ manifold.  Each such
coordinate identification, associated with a
relation~\eqref{phiw}, leads to an embedding ${\rm SO}(4)\subset
G_2$. Indeed, define the action $g_\L$ of any $\L\in{\rm SO}(4)$
on ${\bf x}$ as
\begin{equation}
 g_\L ({\bf x}) = \left(\begin{array}{c}\L\bzeta\\O_+(\L )\bxi\end{array}\right)\; .
\end{equation}
It is then clear from the definition~\eqref{Rpm} of the right-handed vector
representation $O_+$, and Eqs.~\eqref{wb1} and \eqref{phiw}, that
the three-form $\bar{\vf}$ is invariant under $g_\L$ and, hence, $g_\L\in G_2$.

\vspace{0.7cm}

Let $X$ be a seven-dimensional oriented manifold. A $G_2$ structure on $X$
is defined by a smooth three-form $\vf$ which is isomorphic to the
``flat'' $G_2$ structure $\bar{\vf}$ given in Eq.~\eqref{phib}.
The isomorphism induces a metric $g$ on $X$ that is referred to as
the metric associated with $\vf$. Given such a $G_2$ structure $\vf$ the
associated metric can be explicitly computed. Defining
\begin{equation}
 \g_{AB}=\frac{1}{144}\vf_{ACD}\,\vf_{BEF}\,\vf_{GHI}\,\hat{\e}^{CDEFGHI}
 \label{gammadef}
\end{equation}
with the ``pure-number'' Levi-Civita pseudo-tensor $\hat{\e}$, the associated
metric $g$ is given by
\begin{equation}
 g_{AB} = {\rm det}(\g )^{-1/9}\g_{AB}\; ,\qquad \sqrt{{\rm det}(g)} =
          {\rm det}(\g )^{1/9}\; .
 \label{gdef}
\end{equation}
A number of useful properties of $\vf$ can be directly deduced from its
flat counterpart $\bar{\vf}$. For example, one has
\begin{equation}
 \vf_{ABC}\vf^{ABC}=42
\end{equation}
where the indices have been raised with the associated metric $g$. The volume
of the manifold $X$ measured with the metric $g$ can then be written as
\begin{equation}
 {\rm vol}(X) = \int_X\sqrt{{\rm det}(g)}\, d^7x=\frac{1}{7}
                \int_X\vf\wedge\T (\vf )\; .
\end{equation}
Here, the map $\T$ is defined as
\begin{equation}
 \T (\vf )=\star\vf\; ,
\end{equation}
where the Hodge star is taken with respect to the metric $g$
associated with $\vf$. By virtue of~\eqref{gammadef} and
\eqref{gdef}, $\T$ is a highly non-linear map acting on $G_2$
structures $\vf$.

A $G_2$ structure $\vf$ is said to have vanishing torsion if it is
covariantly constant with respect to the Levi-Civita connection
$\nabla$ induced by the associated metric $g$. This condition is
equivalent to $d\vf =d\T (\vf )=0$, or to $\vf$ being harmonic with
respect to the metric $g$.  It can be shown that the holonomy group of
$X$ with respect to $\nabla$ is a subgroup of $G_2$ if the $G_2$
structure is torsion-free. Then, the associated metric $g$ is
Ricci-flat. If, in addition, the first fundamental group $\p_1 (X)$ is
finite, the holonomy group is precisely $G_2$.

\vspace{0.7cm}

In practise, torsion-free $G_2$ structures have not been constructed
explicitly on compact manifolds. Instead, the construction of compact
manifolds with $G_2$ holonomy in Ref.~\cite{joyce1,joyce2,joyceb}
relies on explicit $G_2$ structures with small torsion. It is then
shown that torsion-free $G_2$ structures exist ``nearby''. An
essential ingredient in this construction are small-torsion $G_2$
structures on $F\equiv U\times T^3$, where $U$ is the blow-up of
$\CM^2/\ZM^2$ as defined in Appendix~\ref{EH} and $T^3=\RM^3/\ZM^3$ is
the standard three-torus. Let us now review these specific $G_2$
structures and derive some of their properties.

Following Ref.~\cite{joyce1}, the $G_2$ structure on $F$ is taken to be
\begin{equation}
 \vf = \sum_a w^a\wedge d\xi^a - d\xi^1\wedge d\xi^2\wedge d\xi^3\; .
 \label{phidef}
\end{equation}
Here the two-form $w_1$ has the structure~\eqref{w1} and the functions $\cF$ and
$\cG$ are taken to be the ones associated with the smoothed Eguchi-Hanson
K\"ahler potential~\eqref{SKEH}, and are explicitly given in Eqs.~\eqref{SF}
and \eqref{SG}.
The two other forms $w_2$ and $w_3$ are defined in Eqs.~\eqref{w2} and
\eqref{w3}, respectively. Further, $\xi^a$, where $a,b,\cdots = 1,2,3$,
are the coordinates of the three-torus $T^3$. Since the two-forms $w^a$
are closed the same is true for $\vf$, that is, we have
\begin{equation}
 d\vf = 0\; .
\end{equation}
For the radial coordinate $u$ in the range $u\in [0,u_0]$ the
forms $w^a$ coincide with the three K\"ahler forms of the
Eguchi-Hanson space. On the other hand, in the range, $u\in [u_1
,\s ]$ the $w^a$ are identical to the three K\"ahler
forms~\eqref{wb1} of flat space and, hence, from Eq.~\eqref{phiw},
$\vf$ is given by the flat $G_2$ structure~\eqref{phib} for a
suitable identification of the coordinates $z^\m$ and $\xi^a$ with
$x^A$. Then in both those regions $\vf$ is actually torsion-free,
that is we have $d\T (\vf)=0$ for $u\leq u_0$ or $u\geq u_1$. In
the collar region $u\in [u_0 ,u_1]$, however, $d\T (\vf )$ is
nonzero and can be shown to be of order $\r^4$ and proportional to
derivatives of the interpolating function $\e$. Hence, for small
blow-up radius $\r$ and a smooth interpolation with a
slowly-varying function $\e$, the $G_2$ structure $\vf$ has small
torsion.

\vspace{0.7cm}

The previous statements can be explicitly verified using the metric $g$
associated with $\vf$. We will now compute this metric using
Eqs.~\eqref{gammadef} and \eqref{gdef}. From Eq.~\eqref{phidef} and
the definition of the two-forms $w^a$, Eqs.~\eqref{w1}, \eqref{w2},
\eqref{w3}, \eqref{SF}, along with the expressions for the Maurer-Cartan
forms $\s^a$ given in Appendix~\ref{EH}, we find the only
non-vanishing components of $\vf$ are given by
\begin{eqnarray}
 \vf_{\m\n a} &=& 2m_a{P^0}_{[\m}{P^a}_{\n ]}+n_a{\e^a}_{bc}{P^b}_\m {P^c}_\n \\
 \vf_{abc} &=& -\e_{abc}
\end{eqnarray}
where
\begin{equation}
 (m_a) = \left(\frac{\cF '}{2u},1,1\right)\; ,\qquad
 (n_a) = \left(\frac{\cF}{u^2},1,1\right)\; .
\end{equation}
We recall that the matrix $P$, defined in Eq.~\eqref{P}, is an element
of ${\rm SO}(4)$, a fact which considerably simplifies the subsequent
calculation. Inserting the above components of $\vf$ into
Eqs.~\eqref{gammadef} and \eqref{gdef} leads, after some algebra, to
the associated metric
\begin{equation}
 ds^2 = g_{AB}dx^adx^B=\cG^{-1/3}\left[\frac{\cF '}{2u}du^2
        +\frac{u\cF '}{2}\s_1^2+\cF (\s_2^2+\s_3^2)\right]
        +\cG^{2/3}d\xi_1^2+\cG^{-1/3}(d\xi_2^2+d\xi_3^2)\; .
 \label{gass}
\end{equation}
We note, that the four-dimensional part of this metric differs from
the smoothed Eguchi-Hanson metric~\eqref{gEH1} by the conformal factor $\cG^{-1/3}$.
This difference is due to the complicated relation~\eqref{gammadef},
\eqref{gdef} between $\vf$ and $g$ and it matters precisely in
the collar region where, from Eq.~\eqref{SG}, $\cG$ is different
from one. For the measure associated with the above metric we find
\begin{equation}
 \sqrt{{\rm det}(g)}=\cG^{1/3}\; . \label{detg}
\end{equation}
Note the power $1/3$ by which this result differs from the measure~\eqref{detK}
for the smoothed Eguchi-Hanson metric. Using the explicit expression for
$\cG$ in Eq.~\eqref{SG} we can now compute the volume to order $\r^6$.
One finds
\begin{equation}
  {\rm vol}(0,\s )=\frac{\p^2}{2}\left(\s^4 - \frac{1}{3}\r^4\right)+\cO (\r^8)\; .
\end{equation}
The second term in this expression can be interpreted
as an effect of the blow-up. Due to the non-trivial exponent
in~\eqref{detg} this term is only $1/3$ of the corresponding term
in Eq.~\eqref{Svol} obtained from the smoothed Eguchi-Hanson
metric. This fact can be directly traced to the cubic nature
of the relation~\eqref{gammadef} between $\vf$ and $g$.

\end{document}